\documentstyle[12pt,epsfig]{article}

\newcommand{\be}{\begin{equation}}
\newcommand{\ee}{\end{equation}}
\newcommand{\bees}{\begin{eqnarray}}
\newcommand{\ees}{\end{eqnarray}}
\newcommand{\ra}{\rightarrow}
\newcommand{\pa}{\partial}

\newcommand{\bs}{\bar{S}}
\newcommand{\bt}{\bar{T}}
\newcommand{\vf}{\varphi}
\newcommand{\dvf}{\dot{\varphi}}
\newcommand{\kal}{K\"{a}hler }
\begin{document}

\baselineskip=17pt
\pagestyle{plain}
\setcounter{page}{1}

\begin{titlepage}

\begin{flushright}
 IFUP-TH 8/99\\
 February 1999\\
\end{flushright}

\vspace{5 mm}

\begin{center}
{\LARGE\bf   Loop corrections  and graceful exit}

\vspace{4mm}

{\LARGE\bf     in  string cosmology}
\end{center}

\vspace{4mm}

\begin{center}
{\large  Stefano Foffa$^a$, Michele Maggiore$^a$ and Riccardo
Sturani$^{b,a}$}\\

\vspace{1cm}

{\em (a) Dipartimento di Fisica, via Buonarroti 2, 
I-56100 Pisa, Italy,  \\ and INFN, sezione di Pisa.\\

\vspace{0.3cm}

(b) Scuola Normale Superiore, piazza dei Cavalieri 7, I-56125 Pisa.}

\end{center}

\vspace{2mm}

\begin{center}
{\large Abstract}
\end{center}
\noindent
We examine the effect of perturbative string loops
on the  cosmological pre-big-bang evolution.
We study loop corrections derived from
heterotic string theory compactified 
on a $Z_N$ orbifold and we consider the effect of the {\em all-order} loop 
corrections to the K\"{a}hler potential and of
the corrections to gravitational couplings, including both
threshold corrections and corrections due to the mixed
K\"{a}hler-gravitational anomaly.
We  find that string loops can  drive the evolution
into the region of the parameter space where a graceful exit is 
in principle possible, and we find solutions that,
in the string frame,  connect smoothly the superinflationary 
pre-big-bang evolution to a phase  where
the curvature and the derivative of the dilaton
are decreasing. 
We also find that at a critical coupling the loop corrections to the \kal
potential induce a ghost-like instability, i.e. the kinetic term of
the dilaton vanishes. This is similar to what happens in
Seiberg-Witten theory and signals the transition to a new regime where
the  light modes in the effective action are different 
and are related to the original ones by S-duality.
In a string context, this means that we enter a D-brane dominated phase.
\end{titlepage}
\newpage

\baselineskip=24pt

\section{Introduction}
Pre-big-bang cosmology has been initially
developed using  the lowest-order
effective action of the bosonic string~\cite{GV}. 
This  allowed to understand
the basic features of this cosmological model: the Universe  starts
 at  weak coupling and low curvature, follows a
superinflationary evolution, and enters a large curvature phase.
As long as we are in the
perturbative regime, a description with the lowest order effective
action is adequate, and can be used to discuss the cosmological
evolution~\cite{GV}, 
the generality of the initial conditions~\cite{ini}, and even to
develop some interesting phenomenological consequences concerning the
generation of primordial gravitational, axionic, 
dilatonic and electromagnetic
backgrounds~\cite{phe}.
In particular, the low frequency part of these spectra is only
sensitive to the low curvature part of the evolution.

At lowest order, the cosmological evolution unavoidably reaches  large 
curvatures and strong coupling and finally runs into a singularity.
At this stage it is necessary to go beyond the lowest order 
effective action
for understanding how string theory cures the 
big-bang singularity, and for understanding
the matching of pre-big-bang cosmology 
with standard Friedmann-Robertson-Walker (FRW) post-big-bang
cosmology (the so-called `graceful exit' problem~[4-12]).
One can imagine two possible scenarios. The first
is that pertubative corrections succeed in turning the regime of 
pre-big-bang accelerated
expansion into a decelerated expansion, and at the same time the
dilaton  is
attracted by the minimum of its non-perturbative potential. 
This should take place before entering into a full strong coupling
regime, so that perturbative results can still be trusted. 

In the second scenario the evolution proceeds toward the
full strongly coupled regime. In this case one must take into 
account that at strong coupling and large curvature
new light states appear
and then the approach based on the
effective supergravity action plus string corrections breaks down.
The light modes are now different and one must  turn to a new
effective action, written in terms of the new relevant degrees of
freedom. In particular, at strong coupling  $D$-branes~\cite{Dbrane} 
are expected to play an important
role, since their mass scales like the inverse of the string coupling,
$\sim 1/g$ and they are copiously produced by gravitational
fields~\cite{MR}
(see also refs.~\cite{conifold,DKPS,LW,BFM} for related approaches to
singularities in string theory).

In this paper we investigate how far one can go within 
the first scenario,  analyzing a variety of perturbative
string loop corrections. 
The motivation comes in part from the
 works~\cite{BM1,BM2}, where the authors discuss
general properties that the loop corrections should have in order to
trigger a successful graceful exit. In particular, they find that
the corrections should have the appropriate sign, so to induce
violation of the null energy condition (NEC), but there should also be a
mechanism that at some stage turns them off, otherwise the continued
violation of the NEC produces an unbounded growth in the curvature and 
dilaton.
In ref.~\cite{BM2} they presented a model with loop corrections chosen
{\em ad hoc}, both in the sign and in their functional form, just for the
purpose of verifying explicitly on a toy model the general results of
ref.~\cite{BM1}, and they found that with appropriate choices 
a complete exit transition is indeed obtained. 
A transition to a regime of decreasing curvature was also obtained,
with a potential chosen {\em ad hoc}, in the second paper of
ref.~\cite{GV}. 
This shows that string loops
can in principle trigger a complete graceful exit, 
but leaves open the question of
whether this actually happens for the corrections derived from at least some
specific compactification of string
theory. In particular, the results of
refs.~\cite{BM1,BM2} show first of all that the corrections 
should have the appropriate sign, and this already is an
interesting point to check against  real string derived
corrections, and furthermore  they should
also have a rather non-trivial functional form that suppresses them at
strong coupling.

String loop corrections have been much studied in the
literature, especially for 
$Z_N$ orbifold compactifications of the heterotic string [19-34],
and we can therefore ask whether, at least in some
compactification scheme, they 
fulfill the non-trivial properties needed for a graceful exit.
In particular, corrections to the \kal potential are known
at all loops; this will be very important for our analysis, since in
order to follow the cosmological evolution into the strong coupling
regime, a knowledge of the first few terms of the perturbative
expansion is not really sufficient, and one must have at least some
glimpse into the structure at all loops.

After discussing how far one can go within a perturbative approach, we
will be able to shed some light on the question of whether
perturbative string theory is an adequate tool for discussing string
cosmology close to the big-bang singularity, or whether instead
non-perturbative string physics plays a crucial role, as in the
scenario discussed in ref.~\cite{MR}. 

The paper is organized as follows. In sect.~2 we  discuss the
effective action of the heterotic string with $\alpha '$ and loop
corrections for a $Z_N$ orbifold compactification.
In sect.~3 we use these corrections to study 
the cosmological evolution; we will introduce
various corrections one at the time, to understand better their role,
and we will find solutions that, in the string frame, smoothly
interpolate between the pre-big-bang phase and a phase of decelerated
expansion. 
In sect.~4 we examine the limit of validity of the perturbative
approach. We point out that loop
corrections induce an instability in the kinetic term of the dilaton,
that vanishes at a critical coupling. Similarly to what happens in the
Seiberg-Witten model, beyond this value of the
coupling the effective action is more appropriately written in terms
of variables related to the original ones by S-duality. In string
theory this means that
there is the onset of a new regime where the light modes of the
effective action are given by D-branes rather than by the massless
modes of fundamental strings.
Sect.~5 contains the conclusions. 

\section{The effective action with loop corrections}

We consider the effective action of heterotic string theory 
compactified to four dimensions
on a $Z_N$ orbifold, so that one supersymmetry is left in
four dimensions, and we
restrict to the graviton-dilaton-moduli sector.
In a generic orbifold compactification there are
the $(1,2)$ untwisted moduli fields $U_i$ and
the diagonal $(1,1)$ untwisted moduli fields $T_i$ (non-diagonal
moduli are included in the matter fields).
The moduli fields $U_i$, determine the complex structure,
i.e.  the `shape' of the compact space 
(they generalize the standard modular parameter
of the torus);  in various $Z_N$ orbifolds (see e.g. table~1
in ref.~\cite{DKL}) the Hodge number
$h_{1,2}=0$ and therefore  this shape is fixed
and there are no moduli fields $U_i$. There are instead 3 diagonal
moduli fields $T_i$.  We will
neglect the fields $U_i$ and we restrict to a common diagonal  modulus,
$T_i=T$; it determines the overall volume of compact space.

At the fundamental level string theory compactified on orbifolds
is invariant under $T$-duality, which includes
$SL(2,Z)$ transformations of the common modulus $T$ 
\be\label{modT}
T\ra\frac{aT-ib}{icT+d}\, ,
\ee
with $a,b,c,d\in Z$ and $ad-bc=1$. These modular transformations are 
good quantum symmetries, and therefore they must be
exact symmetries also at the level of the loop-corrected
low-energy effective action.
While at tree level the dilaton is inert under modular
transformations, at one-loop the  cancellation of the
modular anomaly requires that the dilaton transforms as~\cite{DFKZ1}
\be\label{mod}
S\ra S+2\kappa \log (icT+d)\, ,
\ee
where Re~$S=e^{-\phi}$ and $\phi$ is the dilaton field;
$\kappa$ is a positive constant of order one
which depends on the coefficient of the anomaly, see below. Then
one easily verifies that $ S+\bs +2\kappa \log (T+\bt )$ is modular
invariant.
We can therefore introduce a one-loop corrected modular invariant
coupling $g_0^2$ from~\cite{DFKZ1,DFKZ2}
\be\label{g0}
\frac{1}{g_0^2}=\frac{1}{2} (S+\bs )+\kappa \log (T+\bt )
=e^{-\phi}+\kappa\sigma\, ,
\ee
where we have defined the 
field $\sigma$ from ${\rm Re} T=(1/2)e^{\sigma}$ (the factor $1/2$ is
not conventional but we found it convenient). 
Loop corrections to the effective
action  can be computed directly as an expansion in
terms of the modular invariant coupling $g_0^2$~\cite{DFKZ2}. 
As we see from eq.~(\ref{g0}),
\be\label{g0b}
g_0^2=\frac{e^{\phi}}{1+\kappa\sigma e^{\phi}},
\ee
and therefore an expansion in $g_0^2$ 
provides a resummation and a reorganization of
the  expansion in $e^{\phi}$. Note in particular that even when $e^{\phi}$
is large, the expansion in $g_0^2$ is still under control if
$\kappa\sigma\gg 1$, i.e. if $\kappa\log V\gg 1$, where $V$ is the
volume of compact space in string units.

We include in the action terms with two derivatives and
terms with four derivatives, i.e. $O(\alpha ')$ corrections to the
leading term. For both the two- and four-derivatives terms we include
modular invariant loop corrections.
We discuss separately the two- and four-derivatives terms in
subsections 2.1 and 2.2.

\subsection{Terms with two derivatives}
In the Einstein frame, where the gravitational term has the canonical
Einstein-Hilbert form, the action for the metric-dilaton-modulus
system compactified to four dimensions is 
\be\label{act0}
{\cal S}_0^E=\frac{1}{\alpha '}\int d^4x \sqrt{-g}\left[
\frac{1}{2}R-K_{i}^{j}\partial_{\mu}z^i\partial^{\mu}\bar{z}_j
\right]\, .
\ee
Here $z^i=(S,T)$, $K_i^j=d^2K/dz^id\bar{z}_j$ and $K$ is the
K\"{a}hler potential. We use 
the conventions $\eta_{\mu\nu}=(-,+,+,+)$,
${R^{\mu}}_{\nu\rho\sigma}=\partial_{\rho}\Gamma^{\mu}_{\nu\sigma}+\ldots$
and $R_{\nu\sigma}={R^{\rho}}_{\nu\rho\sigma}$. The superscript $E$
reminds that this action is written in the Einstein frame.
The Einstein-Hilbert term $\sqrt{-g}R$ is not renormalized at
one-loop~\cite{AGN1} and  this
non-renormalization theorem persists to all orders in perturbation
theory around any heterotic string ground state with at
least $N=1$ space-time supersymmetry~\cite{KKPR2}.
At tree level the \kal potential is
$K_{\rm tree}=-\log ( S+\bs )-3\log (T+\bt )$. It does renormalize,
and at one loop, for heterotic string compactified
on a $Z_N$ orbifold, becomes~\cite{DFKZ1}
\be\label{K0a}
K_{\rm 1-loop}=-\log\left( S+\bs +2\kappa \log (T+\bt )\right)
-3\log (T+\bt )\, ,
\ee
where $\kappa =3\delta^{GS}/(8\pi^2)$. For instance
for $Z_3$ orbifolds $\delta^{GS}=C(E_8)/2=15$, where $C(E_8)$ is
the quadratic Casimir of $E_8$, and therefore $\kappa\simeq 0.57$.
Eq.~(\ref{K0a}) holds at one-loop,
i.e. at first order in an expansion in $1/(S+\bs )$. 
However, the all-orders resummation implicit in eq.~(\ref{K0a}) is 
dictated by the fact that under the modular transformations
given by eqs.~(\ref{modT},\ref{mod}) the combination 
$ S+\bs +2\kappa \log (T+\bt )$ is  invariant, and therefore 
respects the $T$-duality symmetry of the underlying string theory.
In terms of the modular
invariant coupling $g_0^2$ defined in eq.~(\ref{g0})
we can write eq.~(\ref{K0a}) as
\be\label{K0}
K_{\rm 1-loop}=\log\left(\frac{g_0^2}{2}\right)-3\log (T+\bt )\, .
\ee
Eq.~(\ref{K0}) is the leading term of an expansion in  $g_0^2$
of the all-order K\"{a}hler potential. Indeed, 
the K\"{a}hler potential has  been computed at {\em all
perturbative orders}  in ref.~\cite{DFKZ2}, under the
assumption that no 
dilaton dependent corrections other than the anomaly term
are generated in perturbation theory. One defines implicitly
the all-loop corrected coupling $g^2$ from
\be\label{g}
\frac{1}{g^2}=\frac{1}{g_0^2}
+\frac{2\kappa}{3}\log (\frac{1}{g^2}) +{\rm const.}\,\, .
\ee
The all-loop corrected K\"{a}hler potential then reads~\cite{DFKZ2}
\be\label{K}
K=\log\left[\frac{g^2}{2}\left( 1+\frac{\kappa}{3}g^2\right)^{-3}
\right]-3\log (T+\bt )\, ,
\ee
and  at $g_0^2\ll 1$ it reduces to eq.~(\ref{K0}) plus 
terms $O(g_0^2\log g_0^2)$.

The coupling $g^2$ is just the effective gauge coupling that
multiplies the term $F_{\mu\nu}^2$ after taking into account
loop corrections~\cite{DFKZ2}
and, as we will see in the next section, 
it also multiplies the four derivative term. So the dilaton enters the
action only through $g^2$.
We therefore  define a new  field $\varphi$ from
\be
g^2=e^{\varphi}
\ee
and we will treat it as our fundamental dilaton field. Therefore our 
loop-corrected two-derivative action
is given by eq.~(\ref{act0}) 
where now $z_i=(S',T)$, Re~$S'=e^{-\varphi}=g^{-2}$ and $K$ is given
by eq.~(\ref{K}).

In the following, it will be convenient to work in the string
frame. In four dimensions, we define the string frame metric
$g_{\mu\nu}^S$ in terms of the metric in
the Einstein frame  $g_{\mu\nu}^E$ from
$g_{\mu\nu}^E=g_{\mu\nu}^S e^{-\varphi}$ (in four dimensions
the dilaton  is  the
same in the two frames). Note that we use $\varphi$ rather than $\phi$
to transform between the two frames. At lowest order in $e^{\phi}$ of
course this reduces to the standard definition, but beyond one-loop
the definition in terms of $\varphi$ is more convenient.

Writing explicitly
the kinetic terms of the  dilaton and modulus field, (and omitting the
superscript $S$ for string frame quantities) the 
loop-corrected two-derivative
action in the string frame reads
\be\label{2der}
{\cal S}_{0} =\frac{1}{2\alpha '}\int d^4x \sqrt{-g} e^{-\varphi}
\left[
R+\left( 1+e^{\varphi}G(\varphi )
\right)\pa_{\mu}\varphi\pa^{\mu}\varphi
-\frac{3}{2}\pa_{\mu}\sigma\pa^{\mu}\sigma\right]\, ,
\ee
where we have defined
\be\label{G}
G(\varphi )=\left(\frac{3\kappa }{2}\right) \frac{6+\kappa
e^{\varphi}}{(3+\kappa e^{\varphi})^2}\, .
\ee

\subsection{Terms with four derivatives}

\subsubsection{The  four-derivatives term at tree level}

For the four derivative term we find convenient to work directly in
the string frame. 
At tree-level it can be written as~\cite{CFMP,MT}
\be\label{S1}
({\cal S}_1)_{\rm tree}=\frac{1}{2\alpha '}\int d^4x\sqrt{-g}\,
\left(\frac{k\alpha '}{4}\right)\frac{S+\bs }{2}
\left[
R_{\mu\nu\rho\sigma}R^{\mu\nu\rho\sigma}+bR_{\mu\nu}R^{\mu\nu}+c
R^2\right] 
\ee
and, for the heterotic
string, $k=1/2$ (still, we display $k$ explicitly in our equations to
make easier the comparison with ref.~\cite{GMV}; note however that we
use the opposite metric signature compared to ref.~\cite{GMV}).
The effective action can be obtained requiring that it reproduces the
string amplitudes. This procedure fixes the coefficient of
$R_{\mu\nu\rho\sigma}R^{\mu\nu\rho\sigma}$; however, 
the coefficients $b,c$ cannot be determined from the comparison with
(on-shell) string amplitudes; this can be understood, for instance, 
showing that the coefficients $b,c$ do not enter in the computation of
amplitudes with three on-shell gravitons, while in the computations
of four-graviton amplitudes they cancel between the contact and
exchange graphs~\cite{FOTW}.

A related source of ambiguity appears when one truncates  the
perturbative expansion in $\alpha '$  at any finite order. In fact,
suppose that we choose somehow a value for $b,c$ (two natural
possibilities are either $b=c=0$ or $b=-4,c=1$, which forms the
Gauss-Bonnet combination). Still, we can always
perform a field redefinition that mixes different orders in 
$\alpha '$, e.g. 
 $g_{\mu\nu}\ra g_{\mu\nu}+\alpha '
(a_1R_{\mu\nu}+a_2\pa_{\mu}\phi \pa_{\nu}\phi +\ldots )$, and
 $\phi\ra\phi +\alpha 'b_1R+\ldots$.
This would not change the physics if one would be able to
include all orders in the $\alpha '$ expansion, but it does make a
difference if we truncate at a finite order in $\alpha '$. For
instance, ---retaining only two- and four-derivative terms--- this
redefinition, applied to the two-derivative terms, generates new four
derivative term, while the four-derivatives terms generates
six-derivative terms which are truncated if we work at order
$\alpha '$. The terms which are generated at the four derivative level 
by a generic field redefinitions are
just the terms $R_{\mu\nu}R^{\mu\nu}, R^2$, plus other operators of
the same order involving the dilaton,
like $(\pa\phi )^4, R^{\mu\nu}\pa_{\mu}\phi\pa_{\nu}\phi$, etc. 
This is another way to understand
why the coeffcients of these terms cannot be fixed by the computation
of string amplitudes. 
The term $R_{\mu\nu\rho\sigma}^2$ is instead unaffected by field
redefinitions, consistently with the fact that its coefficient is
fixed by the comparison with  string amplitudes.
The study of the cosmological evolution using
any specific action truncated at order $\alpha '$ should then be considered
as only indicative of the possible cosmological behaviours. With
suitable choices, one finds that the lowest-order pre-big-bang
solutions are indeed regularized by the inclusion of $\alpha '$
corrections, and instead of running into the singularity, they
are now matched (in the string frame) to a phase of De~Sitter
expansion with linearly growing dilaton~\cite{GMV}. The effect on the
cosmological evolution of these ambiguities have been studied in
refs.~\cite{GMV,MM,BM3}. Here
will take the point of view that, independently of these ambiguities,
a solution that, thanks to suitably chosen $\alpha '$ corrections,
 approaches asymptotically  a De~Sitter phase with linear dilaton is a
simple way to model a regularizing effect, which can have a different
and deeper physical motivation: for instance, in ref.~\cite{FMS1} we
found that a similar transition to a De~Sitter phase with linearly
growing dilaton is driven by the formation of a gravitino-dilatino
condensate. This mechanism is independent of the ambiguities
discussed, and only depends on the dynamical assumption that such a
condensate forms.

Our choice for the form of the tree-level-four derivative term is the same
used in refs.~\cite{GMV,BM2}: 
\be\label{S1tree}
({\cal S}_1)_{\rm tree}=\frac{1}{2\alpha '}
\left(\frac{k\alpha '}{4}\right)
\int d^4x\sqrt{-g}\, e^{-\phi}
\left[
R_{\rm GB}^2-(\partial\phi )^4
\right]\, .
\ee
In ref.~\cite{GMV} this form was obtained  
setting $b=c=0$ in eq.~(\ref{S1}) 
and then performing a field redefinition that generates $R_{\mu\nu}^2$
and $R^2$ with the right coefficients to produce
the Gauss-Bonnet combination
$R_{GB}^2=R_{\mu\nu\rho\sigma}^2-4R_{\mu\nu}^2+R^2$, and 
this also generates the term $(\pa\phi )^4$.

Actually, our two-derivative action differs from that used
in~\cite{GMV} already at tree level, because we have included the field
$\sigma$. The field redefinition that generates the Gauss-Bonnet term
will also generate a number of four-derivative terms which depends
also on $\partial{\sigma}$. We will neglect these terms because, on the one hand
they make the action much more complicated, and on the other hand they
are basically irrelevant to the dynamics, as will be clear from the
results of sect.~3 and as we have checked on some examples. 
Actually, because of the ambiguities intrinsic in
a truncation at finite order in $\alpha '$, it is not very meaningful
to insist on any specific form of the action, and it is more important
to look for properties shared at least by a large class of actions
compatible with string theory.

\subsubsection{The loop-corrected four-derivatives term}

Let us first recall what happens in the slightly simpler case
of a gauge coupling $g_a$, where the index $a$ 
refers to the gauge group under consideration. 
The coupling $1/g_a^2$ is identified as
the coefficient of $(1/4)F_{\mu\nu}^aF^{a\mu\nu}$. At tree level,
this term only appears when we expand in components the superfield
expression $(-1/2)f_a(S)W^aW^a$, where $W^a$ is the chiral superfield
containing $F_{\mu\nu}^a$; ${f_a}(S)_{\rm tree}=S$ 
is independent of the gauge group, and 
$1/g_a^2={\rm Re} S$. At one-loop $f_a(S)$ gets a moduli-dependent
renormalization [19-34]
\be
{f_a}(S)_{\rm 1-loop}=S+\Delta_a(T,\bt )\, ,
\ee
For orbifolds with no $N=2$ subsector, such as $Z_3$ and $Z_7$,
$\Delta_a (T,\bt )=\delta_a$ is a moduli-independent constant, and
there is no moduli-dependent
one-loop correction. Beyond one-loop, $f_a(S)$ is 
protected by a non-renormalization theorem~\cite{SV}. 

Furthermore, at one loop  a contribution to $F_{\mu\nu}F^{\mu\nu}$ comes
from the anomaly:
in fact, since  the fermions in the supergravity-matter action are
chiral, the tree level effective action
 leads, through  triangle graphs,
to  one-loop anomalies. The type of anomaly depends on the 
connections  attached to the vertices.
In particular, because of K\"{a}hler symmetry, chiral fermions are
coupled to a $U_K(1)$ K\"{a}hler connection, which is a non-propagating
composite field. Modular
transformations are a subset of K\"{a}hler transformations and
therefore, if the theory is K\"{a}hler invariant, it is also invariant
under modular transformations.
Considering a triangle graph with one K\"{a}hler connection
and two gauge bosons attached at the vertices, we get a mixed 
K\"{a}hler-gauge anomaly,
proportional to $F_{\mu\nu}^a\tilde{F}^{a\mu\nu}$. 
The anomaly can be represented in
the effective theory with a non-local term 
whose (local) variation reproduces the anomaly.
Because of supersymmetry, this effective non-local term, when expanded
in component fields,
together with $F_{\mu\nu}^a\tilde{F}^{a\mu\nu}$ also contains  a term
proportional to $F_{\mu\nu}^aF^{a\mu\nu}$ which, restricting to a
common modulus,  reads~\cite{DFKZ1}
\be\label{nonlocal}
-\left(\frac{\kappa}{3}\right)\left[ \hat{K}(T,\bt )-
2\Box^{-1}\frac{\pa^2\hat{K}}{\pa T\pa \bt} 
\pa^{\mu}T\pa_{\mu}\bt\right] \frac{1}{4}F_{\mu\nu}^a F^{a\mu\nu}\, .
\ee
with $\hat{K}(T,\bt )=-3\log (T+\bt )$.\footnote{A direct
derivation of the term $\sim F_{\mu\nu}F^{\mu\nu}$ from 
Feynman graphs is
quite subtle since, in the formulation of supergravity with the
canonically normalized Einstein term, bosonic currents
do not couple to the K\"{a}hler connection and therefore bosons
cannot run into the triangle
graph; this apparent puzzle is discussed in ref.~\cite{CO2}.} 
In the limit of constant $T$
this term becomes local, and gives an additional  moduli-dependent
one-loop contribution to the gauge coupling $g_a^2$. Therefore, 
specializing for the moment to a $Z_3$ or $Z_7$ orbifold, 
\be\label{gauge}
\frac{1}{g_a^2}=\frac{S+\bs }{2}+
\kappa\log (T+\bar{T}) +\delta_a\, .
\ee
The variation of the term $\kappa\log (T+\bar{T})$ under modular
transformation is just the anomaly, and
the requirement of anomaly cancellation imposes the transformation
law, eq.~(\ref{mod}), 
on the dilaton field. Note that in the case of $Z_3$ and $Z_7$
orbifolds the moduli-dependent part of $\Delta_a$ vanishes, and
therefore it cannot contribute to the cancellation of the anomaly; the
cancellation comes entirely from the variation of $S$; 
so,  in this case the anomaly must be 
independent of the gauge group, as indeed checked in ref.~\cite{DFKZ1}.
Using the linear multiplet formalism~\cite{DQQ}, 
one realizes that this cancellation mechanism is just a
four-dimensional version of the Green-Schwarz mechanism.

At all loops, $g_a^2$ in eq.~(\ref{gauge}) is replaced by definition
with the all-loop corrected effective coupling,
which is the quantity that appears in the all-loop corrected \kal
potential, eq.~(\ref{K}) \cite{DFKZ2}. 

Let us now discuss
four-derivative gravitational couplings, which are the ones relevant for
our analisys. 
In this case we have  three couplings, multiplying 
$R_{\mu\nu\rho\sigma}^2,R_{\mu\nu}^2,R^2$.
The one-loop renormalization of these coupling has been studied in 
refs.~[26-31].
The situation is similar to the
case of gauge couplings, and the contributions again come from the
threshold corrections and from the anomaly (in this case, a mixed
\kal -gravitational anomaly, i.e. a triangle graph with one \kal
connection and two spin connections attached to the vertices).
There are however some complications: first of all,
only one combination of these
operators, corresponding to the the square of the Weyl tensor, i.e. to 
$R_{\mu\nu\rho\sigma}^2-2R^{\mu\nu}R_{\mu\nu}+(1/3)R^2$, is obtained
from a holomorfic function~\cite{CLO}, 
and the other two independent combinations are not
protected by a non-renormalization theorem. Since, in terms of
superfields, the other two combinations that can be formed depend only
on $R_{\mu\nu}^2$ and $R^2$, this means  that
$R_{\mu\nu\rho\sigma}^2$ or $R_{\rm GB}^2$ are protected, while naked
$R_{\mu\nu}^2, R^2$ terms are not.
Furthermore, the ambiguity
due to field redefinitions and truncation at order $\alpha '$ that we
have discussed at tree level, persists of course at one and higher
loops, so that terms like $(\pa\varphi
)^4,R^{\mu\nu}\pa_{\mu}\vf\pa_{\nu}\vf$, etc., as well as term that
depend on $G(\vf )$, can be generated with
field redefinitions. It is then clear that the most general action is very
complicated. We have  chosen to focus on the loop correction
to the same operator that we have considered at tree level, i.e. to
the combination $R_{\rm GB}^{2}-(\pa\vf )^4$, neglecting naked
$R_{\mu\nu}^2,R^2, (\pa\vf )^4$ terms, as well as other terms that can
be generated by field redefinitions. It is of course possible to
extend our analysis including other operators, but we believe that our
choice is sufficient to illustrate the general role of loop corrections, while 
at the same time the action retains a sufficiently simple form, and in
particular the equations of motion remain of second order.
Our four-derivative action is therefore ${\cal S}_1+{\cal S}_{\rm  nl}$, 
where
\be
{\cal S}_1=\frac{1}{2\alpha '}
\left(\frac{k\alpha '}{4}\right)
\int d^4x\sqrt{-g}\, \left( e^{-\varphi}+\Delta(\sigma)\right)
\left[
R_{\rm GB}^2
-(\partial\varphi )^4
\right]
\label{sumup}
\ee
and ${\cal S}_{\rm nl}$ is the non-local contribution from the anomaly,
\be\label{nl}
{\cal S}_{\rm nl}= 
\frac{1}{2\alpha '}
\left(\frac{k\alpha '}{4}\right)\frac{2\kappa}{3}
\int d^4x\sqrt{-g}\,
R_{\rm GB}^2
\Box^{-1}\left(\frac{\pa^2\hat{K}}{\pa T\pa \bt} 
\pa^{\mu}T\pa_{\mu}\bt\right)
\, .
\ee
In the following we will neglect the non-local term. However, an effect
of the anomaly is still present, because 
it has also produced the local contribution necessary to 
turn $e^{-\phi}$ into the
modular-invariant combination $e^{-\vf}$.
Threshold corrections produce the function $\Delta$, 
\be\label{Delta}
\Delta (T,\bar{T})=-\frac{\hat{b}_{\rm gr}}{4\pi^2}
\log\left[ (T+\bar{T})|\eta (iT)|^4\right]+\delta_{\rm gr}\, .
\ee
The constant $\delta_{\rm gr}$ depends on the orbifold considered, and
typical values are estimated in ref.~\cite{Che}.
The constant $\hat{b}_{\rm gr}$ is related to the number of chiral,
vector, and spin-$\frac{3}{2}$ massless super-multiplets,
$N_S,N_V,N_{3/2}$ respectively, by~\cite{ART}
\be
\hat{b}_{\rm gr}=\frac{1}{6}(-3N_V+N_S)-\frac{11}{3}(-3+N_{3/2})
\ee
and vanishes for orbifolds with no $N=2$ subsector as $Z_3$ and $Z_7$; 
$\eta( iT)$ is the Dedeknid eta function,
\be
\eta (\tau )=q^{1/12}\prod_{n=1}^{\infty}\left( 1-q^{2n}\right)
\, ,\hspace{20mm}q\equiv e^{i\pi\tau }\, .
\ee
The anomaly produces also a term
$\sim R_{\mu\nu\rho\sigma}\tilde{R}^{\mu\nu\rho\sigma}$. However,
below we
will specialize to a metric of the  FRW form, and in this background
$R_{\mu\nu\rho\sigma}\tilde{R}^{\mu\nu\rho\sigma}$ vanishes
identically (which allows us to look for solutions of the equations of
motion with Im~$S=0$, Im~$T=0$~\cite{ART}).

\section{The cosmological evolution}

We now restrict to an isotropic FRW metric with scale factor 
$a(t)=e^{\beta(t)}$, and Hubble parameter $H=\dot{a}/a=\dot{\beta}$. 
We use $H$ to denote the Hubble parameter in the string
frame. Another 
useful quantity is the Hubble parameter in the Einstein frame, $H_E$, 
related to $H$ and $\varphi$ by
\be
H_E=e^{\varphi /2}\left( H-\frac{1}{2}\dvf\right)\, .
\ee
(Recall that we use $\varphi$ rather than $\phi$ 
to move from the Einstein to the string frame).
The shifted dilaton $\bar{\varphi}$ is defined by
$\dot{\bar{\varphi}}\equiv \dvf -3 H$. In the numerical analysis we
will use units $k\alpha '=1$.

At lowest order in both the $\alpha '$ and the loop expansion
the  solutions of the effective action come in pairs. For
$\dot{\sigma}=0$, they satisfy
$\dot{\varphi}=(3\pm \sqrt{3})H$, or
$\dot{\bar{\varphi}}=\pm\sqrt{3}H$, 
referred to as $(\pm )$ branches
respectively~\cite{GV}. 
The $(+)$ branch, when $H>0$, describes a Universe that
starts from the low curvature regime and follows an accelerated 
superinflationary expansion. This is called a pre-big-bang type
solution, and is characterized by $\dot{\bar{\varphi}}>0$.
The $(-)$ branch, when $H>0$, 
describes instead a post-big-bang evolution with
decelerated expansion and  $\dot{\bar{\varphi}}<0$.
In the present era we have a FRW decelerated expansion with stabilized
dilaton, $\dvf =0, H>0$, and therefore $\dot{\bar{\varphi}}<0$.
A graceful exit is then obtained if the cosmological
evolution connects the $(+)$ branch with a 
 $(-)$ branch solution. Necessary conditions
for this to happen have been discussed by Brustein and
Madden~\cite{BM1}, and it turns out that a rather non trivial
interplay of different types of corrections is required.
First of all, of course, we must move from the region of parameter
space where $\dot{\bar{\varphi}}>0$ to the region with
$\dot{\bar{\varphi}}<0$. This can be accomplished 
with $\alpha '$ corrections~\cite{GMV}.
Furthermore, it can be shown that on the $(+)$ branch $H_E<0$
(in fact, without corrections, $H_E$ is monotonically decreasing
and  runs toward $-\infty$ at the
singularity) and it is still negative at the branch change, i.e., when 
$\dot{\bar{\varphi}}=0$. To conclude the exit successfully, it is
shown in ref.~\cite{BM1} that
the evolution must proceed to a region where $H_E$
is positive. This is in general non-trivial since at the beginning of
the evolution $H_E$ is negative and decreasing, and therefore a bounce
in $H_E$ is necessary. Thus, some new physical effect must be turn on to
produce this bounce, and it is at this point that loop corrections are
supposed to be crucial. It was found in ref.~\cite{BM2} that indeed a
one-loop correction with the appropriate sign turns the decrease of $H_E$
into a growth and drives the solution into
the region $H_E>0$. However, this growth is unbounded, and the
solution is now driven toward $H_E\ra
+\infty$. Therefore it is necessary that, after we have reached the
region $H_E>0$, some new mechanism
turns on and  kills the effect of the one-loop corrections. This was
modelled in ref.~\cite{BM2} multiplying the loop correction by a
smoothed theta function. Then, after the evolution reaches the region
$H_E>0$ and loops corrections are switched off, 
it becomes  in principle possible to stabilize
the dilaton with a potential.

In this section we study the equations of motion, taking initial
conditions of the pre-big-bang type; we will add various sources of
corrections one at the time, in order to have some understanding of
the role of the various terms, and we will compare with the above
picture.

\subsection{The evolution without loop corrections}
First of all, we examine the behaviour of the system 
including $\alpha '$ corrections, but
without the inclusion of
loop corrections. In this case $\phi =\varphi$ and
our action reads
\be
{\cal S}=\frac{1}{2\alpha '}\int d^4x \sqrt{-g}\, e^{-\varphi}
\left[
R+\pa_{\mu}\varphi\pa^{\mu}\varphi
-\frac{3}{2}\pa_{\mu}\sigma\pa^{\mu}\sigma
+\frac{k\alpha '}{4}\left( R_{\rm GB}^2-(\pa\vf )^4\right)
\right]\, .
\ee

\begin{figure}
\label{fig1}
\centering
\includegraphics[width=0.4\linewidth,angle=-90]{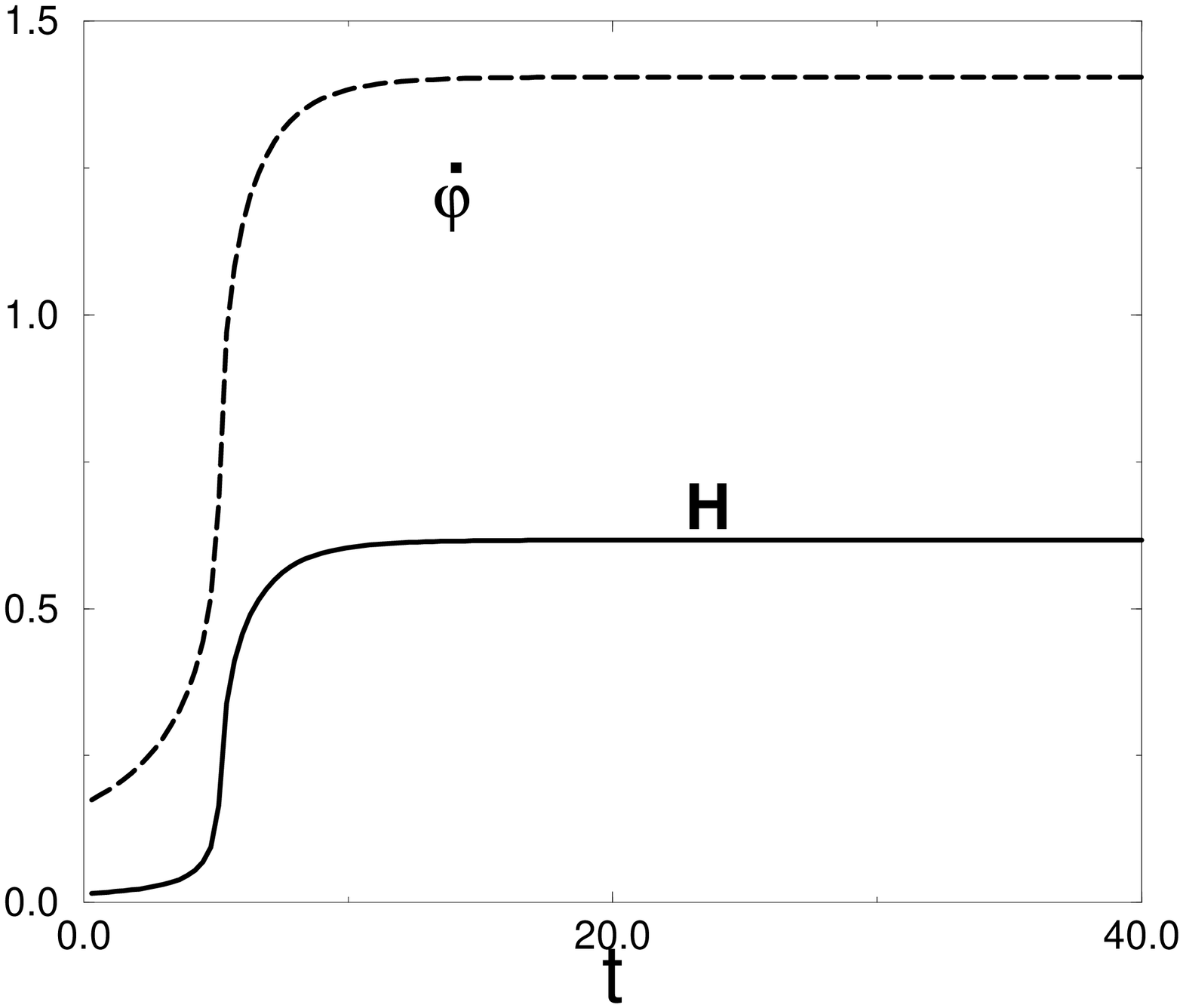}
\includegraphics[width=0.4\linewidth,angle=-90]{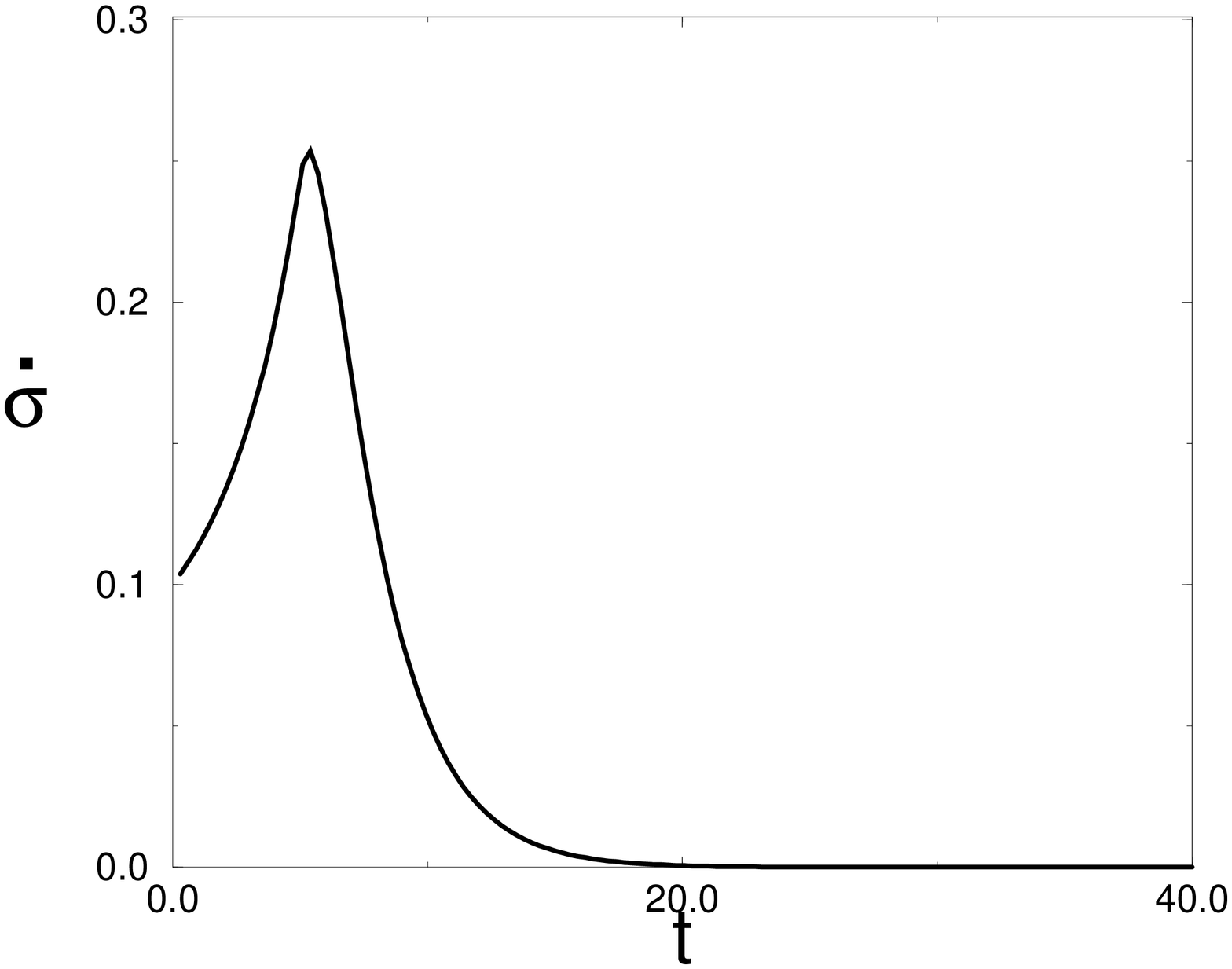}
\mbox{(a)}\mbox{\hspace{7cm}}\mbox{(b)}
\caption{(a) Evolution of $H$, $\dot\varphi$
for the tree-level system.
Initial conditions at $t=0$ are: $H(0)=0.015$, $\varphi (0)=-30$,
$\dot\sigma (0)=0.1$; $\dvf (0)$ is then fixed by the constraint
equation, $\dvf =0.16781\ldots $. 
(b) The evolution of $\dot{\sigma}$.}
\end{figure}

\begin{figure}
\label{fig2}
\centering
\includegraphics[width=0.6\linewidth,angle=-90]{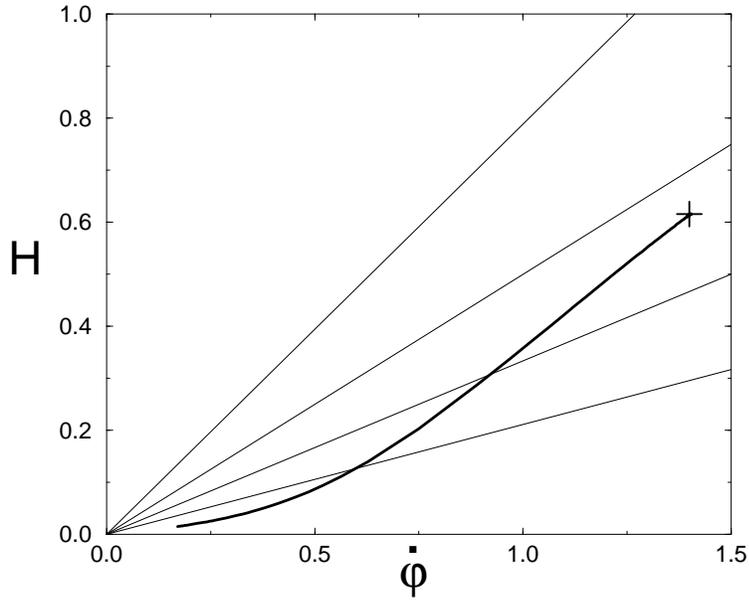}
\caption{The evolution in the $(H,\dot{\varphi})$ plane. The four
lines, in order of increasing slope, are the $(+)$ branch, the branch
change line, the bounce line, and the $(-)$ branch, see the text.}
\end{figure}

If we neglect the modulus field $\sigma$, this action
reduces to that considered in ref. \cite{GMV} and then we know that,
starting from initial conditions of the pre-big-bang type, the solution
has at first the usual pre-big-bang
superinflationary evolution and then, when the
curvature becomes  of order one (in units $k\alpha '=1$), it feels
the effect of the $\alpha '$ corrections. At this stage, instead of running
into the singularity, it is attracted 
towards a fixed point with $H$ and $\dot{\varphi}$ constants. This
picture is not modified by the inclusion of $\sigma$. In fact,
writing also the equation of motion for $\sigma$, 
one immediately sees that there is
an algebraic solution of the equations of motion with
$\dot{\sigma}=0$, and $H,\dvf$ constant and 
the same as in ref.~\cite{GMV}, i.e. $H=0.616\ldots ,\dvf =1.40\ldots$.
The numerical integration, see Figs.~1a,1b, 
shows that this solution is still an attractor of the pre-big-bang
solution. 

For the discussion of the graceful exit, 
it is very convenient to display the solutions also in the
$(H,\dot{\varphi})$ plane, following ref.~\cite{BM2}. In this graph, 
shown in Fig.~2, four lines are of special interest.
In order of increasing slope,
the first line is the
$(+)$ branch of the lowest order solution (more precisely, this line
corresponds to the lowest order solution only in the limit
$\dot{\sigma}=0$, and the deviation of the initial evolution
from it that we see in Fig.~2 is due to a non-vanishing initial value
of $\dot{\sigma}$).
The second  line corresponds to 
branch change, i.e. $\dvf -3H=0$. The third is the line where $H_E=0$, and
as found in~\cite{BM1}, it is necessary that the evolution
crosses also this line to complete the exit.
Finally, we have the 
line representing a $(-)$ branch solution.
We see from Fig.~2 that the lowest order solution
ends up at a fixed point, after
crossing the branch change line, but it is still
in the region $H_E<0$.

The solution shown in this subsection can be considered as the starting point
of our analysis; in the following subsections we will see how the various
loop corrections modify this basic picture.

\subsection{The effect of the loop-corrected K\"{a}hler potential}
To begin our analysis we restrict to a $Z_3$ orbifold, so that
 threshold corrections vanish,
and we also neglect the non-local term. 
The action that we use in this section is therefore
\be
{\cal S}=\frac{1}{2\alpha '}\int d^4x \sqrt{-g} e^{-\varphi}
\left[
R+\left( 1+e^{\varphi}G(\varphi )
\right)\pa_{\mu}\varphi\pa^{\mu}\varphi
-\frac{3}{2}\pa_{\mu}\sigma\pa^{\mu}\sigma
+\frac{k\alpha '}{4}\left( R_{\rm GB}^2-(\pa\vf )^4\right)
\right]
\ee
with
\be
G(\varphi )=\left(\frac{3\kappa}{2}\right) \frac{6+\kappa
e^{\varphi}}{(3+\kappa e^{\varphi})^2}\, .
\ee
We again restrict to isotropic FRW metric and homogeneous fields and
write the equations of motion for the fields $\vf (t),\sigma (t),
\beta (t)$.
Taking the variation with respect to $\sigma$, we get the equation of
motion 
\be\label{sigma}
\frac{d}{dt}\left( e^{3\beta -\vf}\dot{\sigma}\right)=0\, .
\ee
Therefore, if we take as initial
condition $\dot{\sigma}=0$,  $\sigma $ will stay constant.
In this case the non-local term,  eq.~(\ref{nl}),
vanishes at all times, and therefore, for this initial condition, no
approximation is made omitting it.

Before starting with the full numerical integration it is useful
to make contact with the general analisys
of ref.~\cite{BM2}. We therefore restrict to constant $\sigma$ 
(since $\sigma$ was not included in  ref.~\cite{BM2}) and
we  write the equation of motion obtained 
with a variation of the lapse function 
in the form
\be\label{30}
6H^2+\dvf^2-6H\dvf =e^{\vf}(\rho_{\alpha '}+\rho_{q})
\ee
where 
\be\label{rhoa}
\rho_{\alpha '}=(k\alpha ')
e^{-\vf}\left( 6H^3\dvf -\frac{3}{4}\dvf^4\right)\, 
\ee
is the contribution of the $\alpha '$ corrections.  The contribution of
loop corrections is in the function $\rho_q$ which, from our action,
turns out to be
\be\label{rhoq}
\rho_q=-\dvf^2G(\vf )=-\dvf^2\left(\frac{3\kappa}{2}\right)
\frac{6+\kappa e^{\varphi}}{(3+\kappa e^{\varphi})^2}\, .
\ee
In ref.~\cite{BM1} it was found that a graceful exit could be obtained
with a loop correction that gives
$\rho_q=-3 f(\varphi )\dvf^4$, with $f(\varphi )$ a smoothed theta
function going to zero, for large $\varphi$, as
$e^{-16\varphi}$. This form of the correction was just postulated in
ref.~\cite{BM2}, but
comparing it with the string result, eq.~(\ref{rhoq}),
 we find that, 
first of all, the sign comes out right, which is of course
non-trivial. The dependence is $\sim\dvf^2$ rather than $\dvf^4$ since
it comes from a correction to the kinetic term and,
most importantly, its behaviour at large $\varphi$ is different. In fact
$G(\varphi )$ resembles a smoothed theta function, which is also a
non-trivial and encouraging result, but it goes to zero
only as $e^{-\varphi}$, which just compensate the factor $e^{\varphi}$
in eq.~(\ref{30}). We will see from the numerical analysis that this
produces important differences compared to ref.~\cite{BM2}. 

We now turn to the full numerical analysis, we restore
$\sigma$ as a dynamical field and we set $k\alpha'=1$. 
The equations of motion obtained with a variation with respect to
$\vf$ and $\beta$ are, respectively,
\bees
& &-6\dot{H}(1+H^2)+\ddot{\vf}(2+2e^{\vf}G+3\dvf^2)-12H^2
-\frac{3}{2}\dot{\sigma}^2
-\frac{3}{4}\dvf^4 -6H^4+3H\dvf^3
+\nonumber\\
& &+6(1+e^{\vf}G)H\dvf -\dvf^2 (1-e^{\vf}G')=0\, ,\\
& &4\dot{H}(1-H\dvf)-2\ddot{\vf}(1+H^2)+6H^2-4H\dvf+(1-e^{\vf}G)\dvf^2
-\frac{1}{4}\dvf^4-4H^3\dvf+\nonumber\\
& &+2\dvf^2H^2+\frac{3}{2}\dot{\sigma}^2=0\, ,
\ees
and together with eq.~(\ref{sigma}) they determine the evolution of the
system. The variation with respect to the lapse function produces a
constraint of the initial data,
\be
6H^2+\dvf^2-6H\dvf -\frac{3}{2}\dot{\sigma}^2
=e^{\vf}(\rho_{\alpha '}+\rho_{q})
\ee
with $\rho_{\alpha '}$ and $\rho_q$ given in eqs.~(\ref{rhoa},\ref{rhoq}).
The constraint is conserved by the dynamical equations of
motion. We used this conservation as a check of the accuracy of the
integration routine. Typically, the constraint is zero with an accuracy
of $10^{-5}$.

The result of the numerical integration is shown in Figs.~3a,3b. We
see that at first loop corrections are small and $\dvf ,H$ 
are the same as in Fig.~1a. At some stage loop corrections become
important and the solution settles to a new fixed point, again with 
$\dvf$ and $H$ constant. Instead, at least on the scale used,
the evolution  of $\dot{\sigma}$ is
indistinguishable from the case without loop corrections, compare
Figs.~3b and 1b, because $\dot{\sigma}$  is
practically zero when loop corrections become effective. 
The change of regime takes place when the coupling $g^2$ is of order
one, as can be seen from Fig.~4, where we expand the region in time
where loop corrections become important and we plot $H,\dvf$ and the
coupling $g^2$.

\begin{figure}
\label{fig3}
\centering
\includegraphics[width=0.4\linewidth,angle=-90]{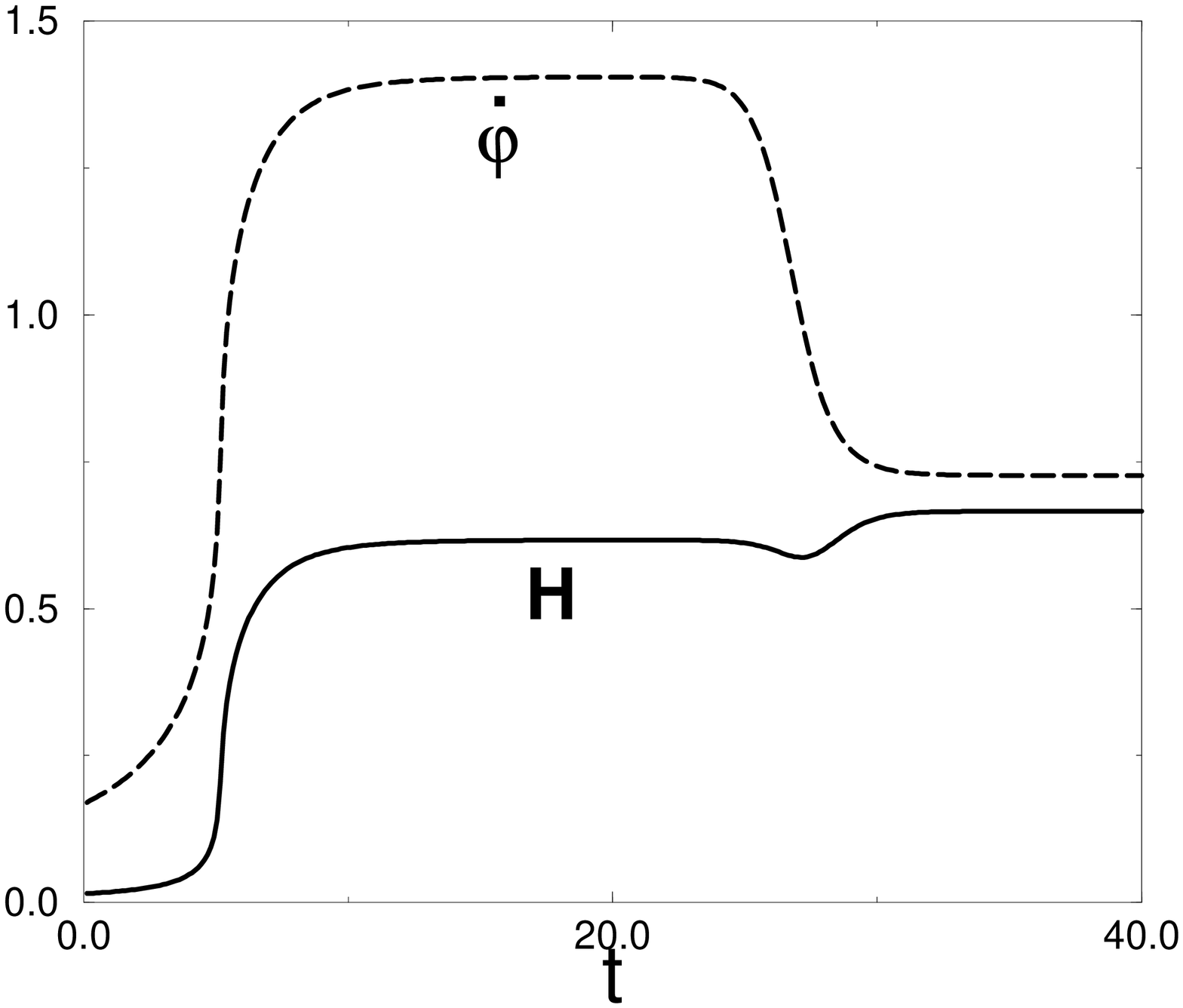}
\includegraphics[width=0.4\linewidth,angle=-90]{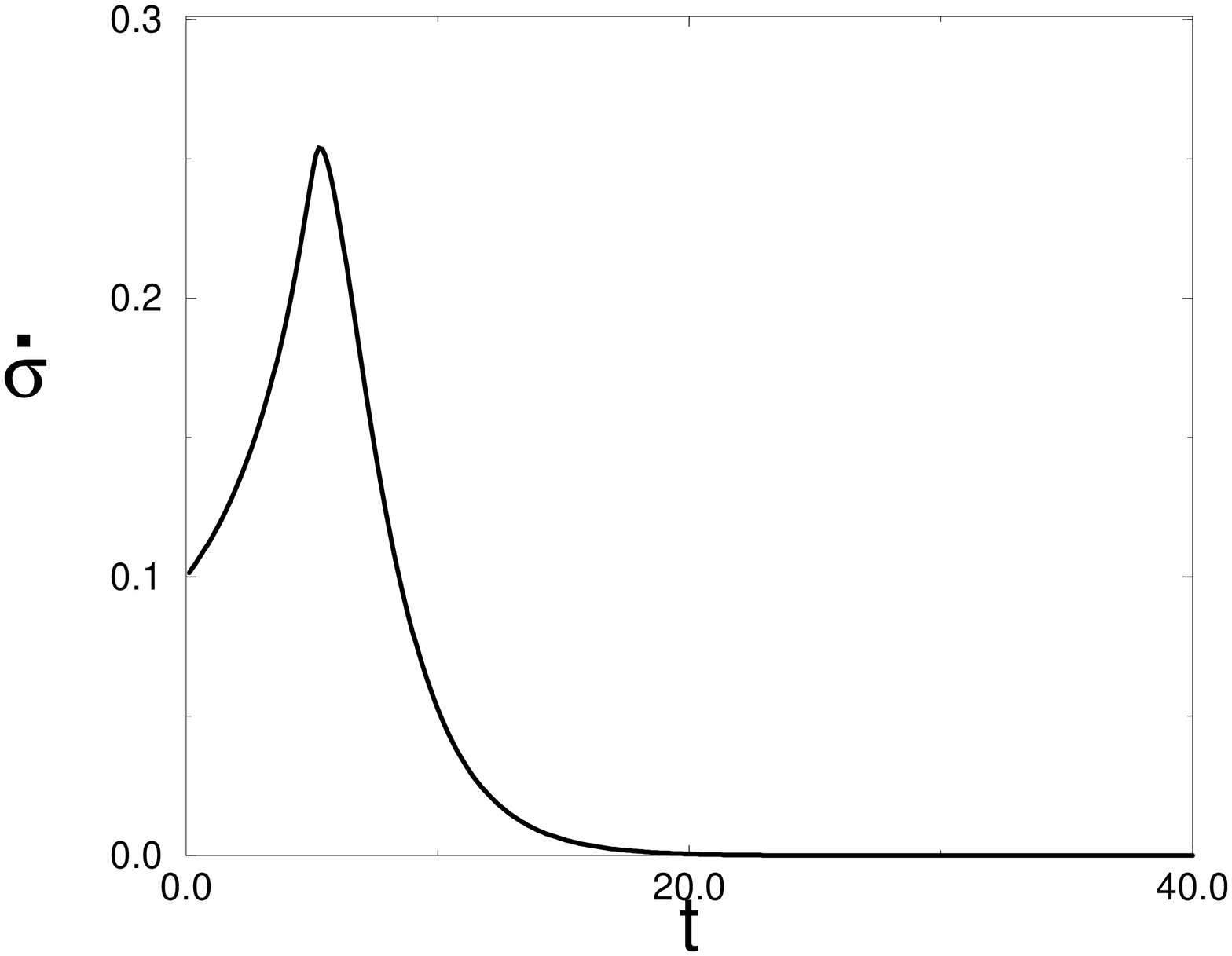}
\mbox{(a)}\mbox{\hspace{7cm}}\mbox{(b)}
\caption{(a) The evolution of $\dvf ,H$ 
including the all-order loop corrections
to the \kal potential; (b) the evolution of $\dot{\sigma}$.
Initial conditions are the same as in the tree-level case and $\kappa=0.57$.}
\end{figure}

\begin{figure}
\label{fig4}
\centering
\includegraphics[width=0.4\linewidth,angle=-90]{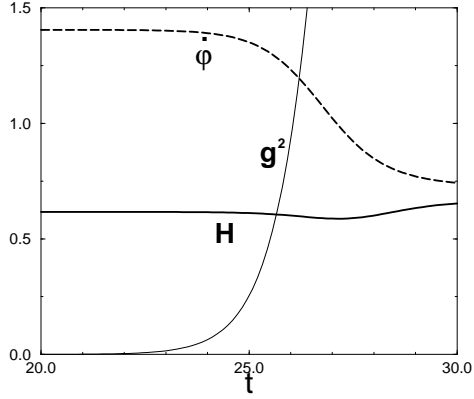}
\caption{$H,\dvf$ and  $g^2=e^{\vf}$ as a function of time. 
Compared to Fig.~3a we have
  expanded the range of $t$ where loop corrections become important.}
\end{figure}

{}From these plots, it might seem that after all
the situation is not so different from the tree level
evolution, because in both cases the solution in the string frame
eventually approaches  a
De~Sitter phase with linearly growing dilaton. An important difference
however is found plotting the solution in the $(H,\dvf)$ plane, see
Fig.~5. We see in fact that the solution has crossed the line $H_E=0$
(and actually even the $(-)$ branch line) and therefore entered the
region of parameter space where a graceful exit is in principle
possible. Plotting the evolution of $H_E$ shows again that the loop
corrections due to the \kal potential  produce a bounce in
$H_E$, see Fig.~6.

\begin{figure}
\label{fig5}
\centering
\includegraphics[width=0.6\linewidth,angle=-90]{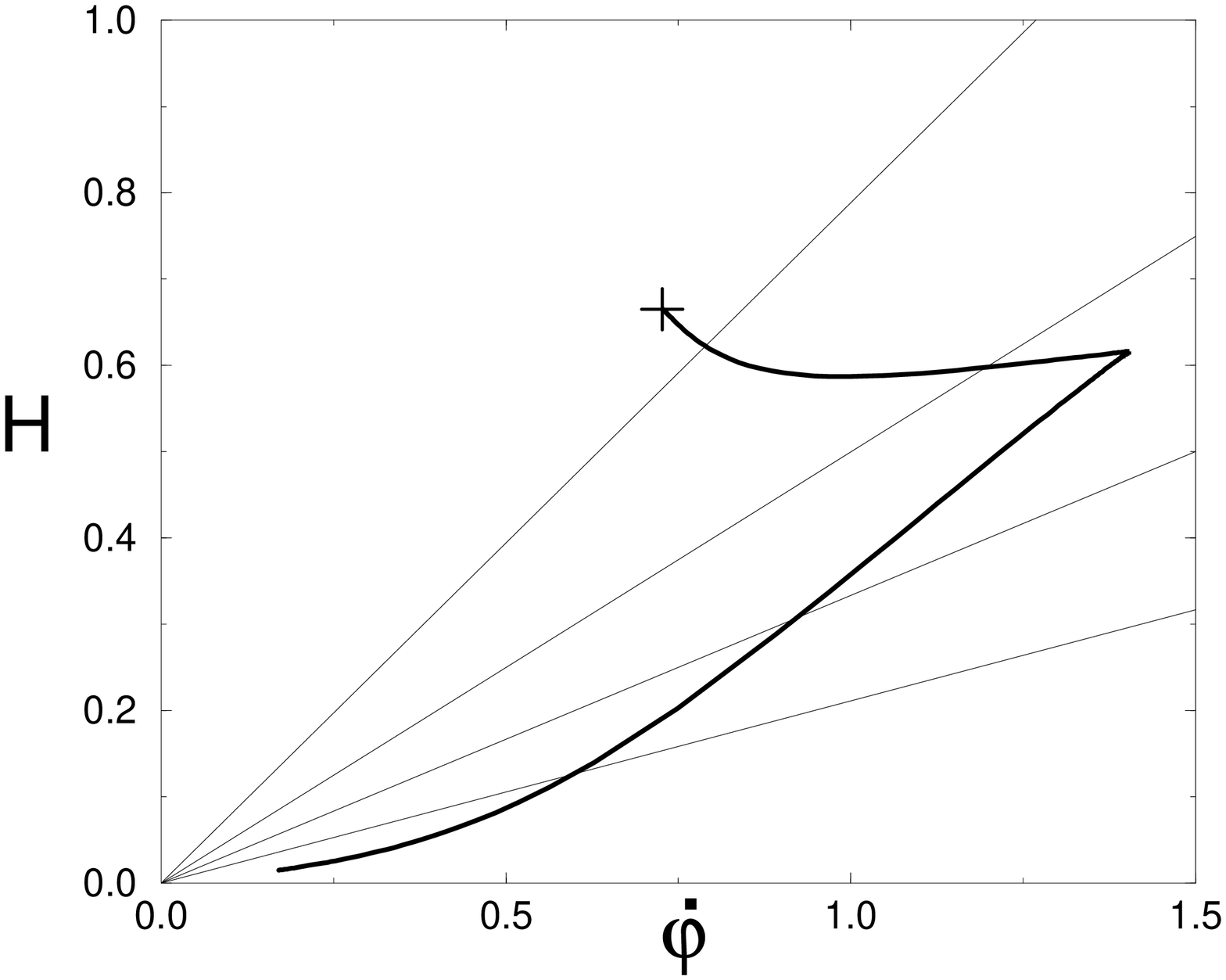}
\caption{The evolution in the $(H,\dvf )$ plane. The straight lines
  are as in fig.~2.}
\end{figure}

\begin{figure}
\label{fig6}
\centering
\includegraphics[width=0.4\linewidth,angle=-90]{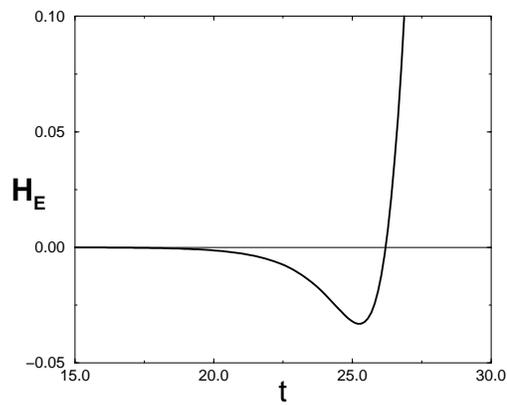}
\caption{$H_E$ as a function of (string frame) cosmic time.} 
\end{figure}

Thus, loop corrections to the \kal potential succeed in doing part of
what loop corrections are expected to do, i.e. they produce a bounce in
$H_E$ and move the solution into the region $H_E>0$. 
However, we also want to
obtain a solution with $H,\dvf$ eventually decreasing and we want to
connect this solution to the $(-)$ branch. We therefore turn to
threshold correction to see if they can produce this effect.

\subsection{The effect of threshold corrections}

We now turn on  the moduli-dependent threshold corrections, so that
the action becomes
\bees
{\cal S}&=&\frac{1}{2\alpha '}\int d^4x \sqrt{-g} \left\{ e^{-\varphi}
\left[
R+\left( 1+e^{\varphi}G(\varphi )
\right)\pa_{\mu}\varphi\pa^{\mu}\varphi
-\frac{3}{2}\pa_{\mu}\sigma\pa^{\mu}\sigma\right]+\right.\nonumber\\
& &\hspace*{15mm}
+\left. \frac{k\alpha '}{4} \left( e^{-\vf}+\Delta (\sigma )\right)
\left[ R_{\rm GB}^2-(\pa\vf )^4\right]
\right\}\, .
\ees
The equations of motion are now (setting again $k\alpha'=1$)
\bees
& &-6\dot{H}(1+H^2)+\ddot{\vf}[2+2e^{\vf}G+3(1+\Delta)\dvf^2]-12H^2
-\frac{3}{2}\dot{\sigma}^2
-\frac{3}{4}\dvf^4 -6H^4+\nonumber\\
& &+[3H(1+\Delta)+\dot{\Delta}]\dvf^3
+6(1+e^{\vf}G)H\dvf -\dvf^2 (1-e^{\vf}G')=0\, ,\\
& &4\dot{H}[1-H(\dvf-\dot{\Delta})]
-2\ddot{\vf}(1+H^2)+6H^2-4H\dvf+(1-e^{\vf}G)\dvf^2
-\frac{1+\Delta}{4}\dvf^4+\nonumber\\
& &-4H^3(\dvf-\dot{\Delta})
+2(\dvf^2+\ddot{\Delta})H^2+\frac{3}{2}\dot{\sigma}^2=0\, ,
\ees
and the constraint on the initial data is
\be
6H^2+\dvf^2-6H\dvf -\frac{3}{2}\dot{\sigma}^2
=e^{\vf}(\rho_{\alpha '}+\rho_{q}+\rho_{q\alpha '})\, ,
\ee
where $\rho_{\alpha '},\rho_q$ are given in
eqs.~(\ref{rhoa},\ref{rhoq}) and
\be
\rho_{q\alpha '}=(k\alpha ') \left
  ( -6\dot{\Delta}H^3
-\frac{3}{4}\Delta \dvf^4\right)\, .
\ee
As initial conditions for $\sigma$ we take a value close to
the self-dual point,
$\sigma (0)\simeq \sigma_{\rm sd}=\log{2}$, that is Re~$T\simeq 1$, and
we take $\dot{\sigma}$ small (consistently with
the fact that the pre-big-bang evolution starts from the flat
perturbative vacuum). We will discuss later the dependence on the
initial conditions.
With these choices, for a generic orbifold
$\Delta(\sigma)$ turns out to be practically constant
during the course of the evolution (and for a $Z_3$ or $Z_7$ orbifold
$\Delta (\sigma )=\delta_{\rm gr}$ is exactly constant) and
its value  is
determined by   $\hat{b}_{\rm gr}$ and $\delta_{\rm gr}$; taking for
instance $\delta_{\rm gr}=0$, we 
have found nonsingular solutions in the range
$\hat{b}_{\rm gr} \in [-20,0)$, which corresponds to
 $\Delta (\sigma_{\rm sd})\in [-0.18,0)$.

The evolution of the system under these conditions is shown in Figs.~7a,7b.
The behaviour of $H,\dvf$ is quite remarkable: threshold corrections
turn the De~Sitter phase with linearly growing dilaton into a phase 
with $H,\dvf$ decreasing! At the same time the modulus $\sigma$, and
therefore the volume of internal space, shows a rather elaborate
dynamics, see Fig.~7b. These figures refer to a $Z_6$ orbifold, for
which $\kappa \simeq 0.19$. The same qualitative behaviour is obtained
for a $Z_3$ orbifold, in which case $\Delta (\sigma )=\delta_{\rm gr}$ is 
exactly constant, and the same results are also obtained for
different, generic, values of $\kappa$.

\begin{figure}
\label{fig7}
\centering
\includegraphics[width=0.6\linewidth,angle=-90]{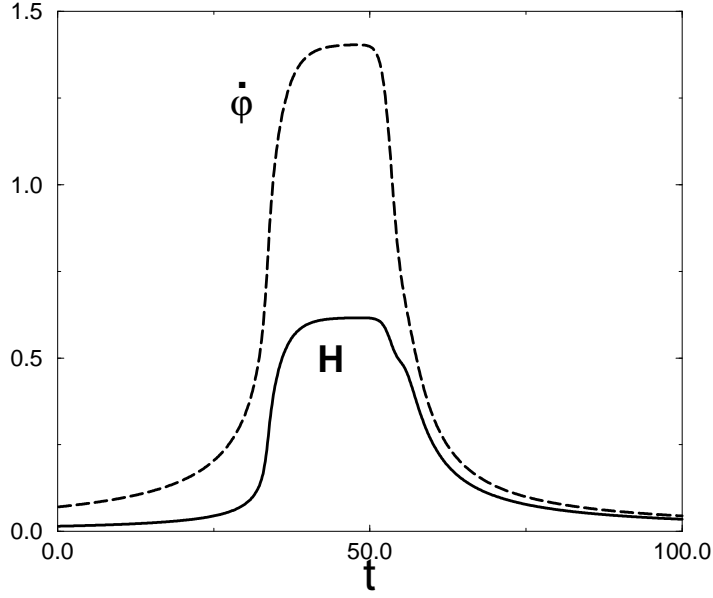}

\mbox{(a)}
\includegraphics[width=0.6\linewidth,angle=-90]{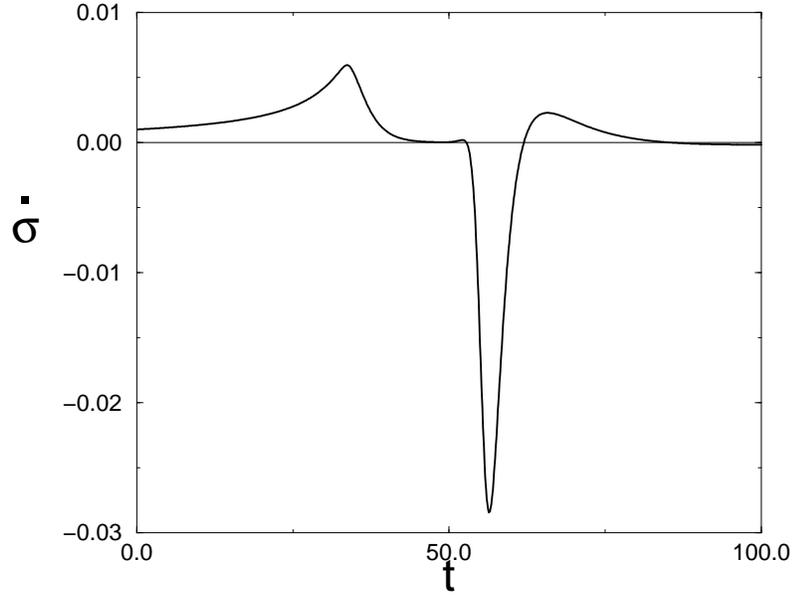}

\mbox{(b)}
\caption{(a) The evolution of $H$ and $\dot{\varphi}$ with
loop corrections to the \kal potential and
threshold correction switched on. The initial conditions are
$H(0)=0.015,\vf (0)=-30$,
$\sigma (0)=0.69$, $\dot\sigma (0)=0.001$, and $\dvf (0)=0.07067\ldots$ is then
fixed by the constraint; the values of the parameters are
$\kappa=0.19$, $\hat{b}_{\rm gr}\simeq-4, \delta_{\rm gr}=0$.
(b) The evolution of $\dot{\sigma}$. }
\end{figure}

The evolution in the $(H,\dvf )$ plane is shown in Fig.~8, and we see
that the solution approaches the $(-)$ branch. From this figure we
also see that the solution approaches at first the tree-level 
fixed point discussed
in sect.~3.1, then
corrections to the \kal potential and the threshold
corrections become important about at the same time, so that after
leaving this fixed point the solution deviates immediately from the
behaviour that it has
in the absence of thresholds corrections, shown in fig.~5, and
it does not get close to the  fixed point marked by a cross in
fig.~5. Instead, if we do not include the corrections to the \kal
potential and we only switch on the threshold corrections, we found
that the solution never crosses the bounce line $H_E=0$, and therefore
the corrections to the \kal potential are really an essential
ingredient of our solution.

Fig.~9 shows instead the evolution of the coupling $g^2$, and we see that
the curvature and the derivative of the dilaton start decreasing when $g^2\sim
1$, so that when the solution is close to the $(-)$ branch we are already
at large $g$, and at this stage non-perturbative effects are expected to become
important. 
We will discuss this point further in sect.~4. 

\begin{figure}
\label{fig8}
\centering
\includegraphics[width=0.6\linewidth,angle=-90]{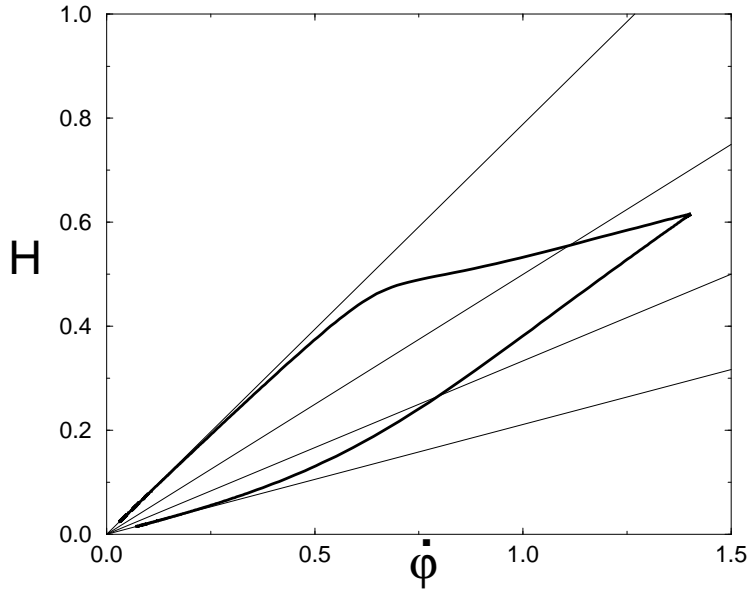}
\caption{The evolution in the $(H, \dvf)$ plane. The straight lines
  are as in fig.~2.}
\end{figure}

\begin{figure}
\label{fig9}
\centering
\includegraphics[width=0.6\linewidth,angle=-90]{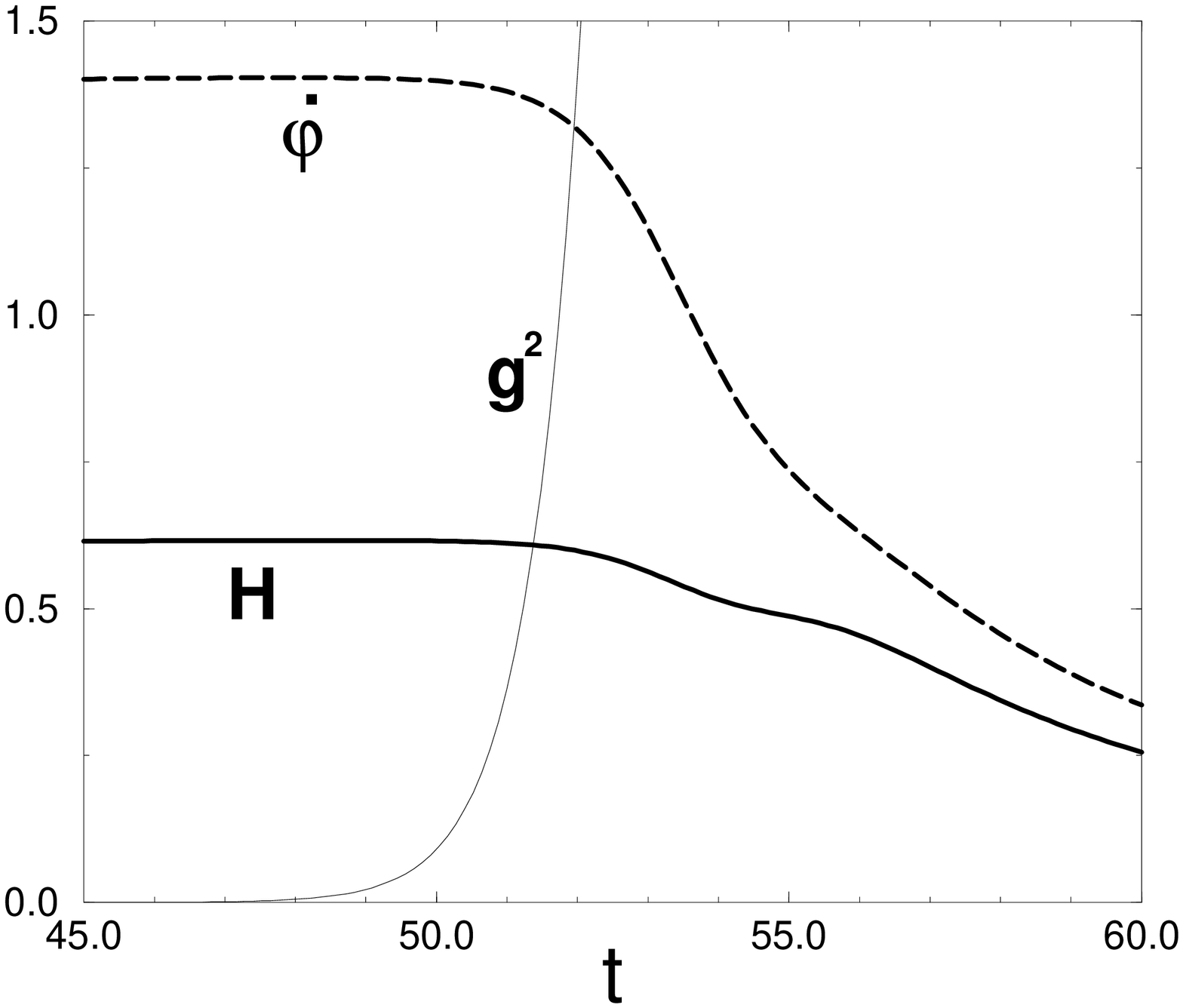}
\caption{$H$, $\dvf$ and $g^2$ against cosmic time.}
\end{figure}

Although it is appropriate to recall at this point that these results are
obtained with some specific choices of action and of initial
conditions, it is certainly interesting to have at least an example
of such a behaviour, with choices well motivated by string theory.
To get some understanding of the dependence on the initial conditions 
we have run the integration routine for many different values of
$\sigma(0)$ and $\dot{\sigma}(0)$. The shaded area in Fig.~10
is the region of the plane $(\sigma (0),\dot{\sigma (0)})$
where the behaviour is qualitatively the same as that
shown above, while for initial conditions  outside the shaded region
the evolution in general runs into a singularity. Considering that
$\sigma$ is at the exponent in Re~$T$, the limitation on $\sigma (0)$ is
not particularly strong, while the required values of $\dot{\sigma}(0)$ are
of the same order as the initial value
of $H$. These initial conditions do not imply therefore any  fine
tuning.

To  have a better understanding of these solutions, it is also useful
to display  the corresponding   Einstein-frame quantities. (We still plot
them against string frame time $t$, but the same qualitative behaviour is
obtained against Einstein frame time $t_E$; the two are related by
$dt=dt_E \exp (\vf /2)$). In Figs.~11a,11b we plot
$\dvf_E=d\vf /dt_E$ and $H_E$. The latter is particularly interesting
and shows that in the Einstein frame our solution approaches
asymptotically a De~Sitter inflation. 
This is of course very
different from the result of ref.~\cite{GMV} or of sect.~3.1, where
De~Sitter inflation takes place in the string frame.

\begin{figure}
\label{fig10}
\centering
\includegraphics[width=0.6\linewidth,angle=-90]{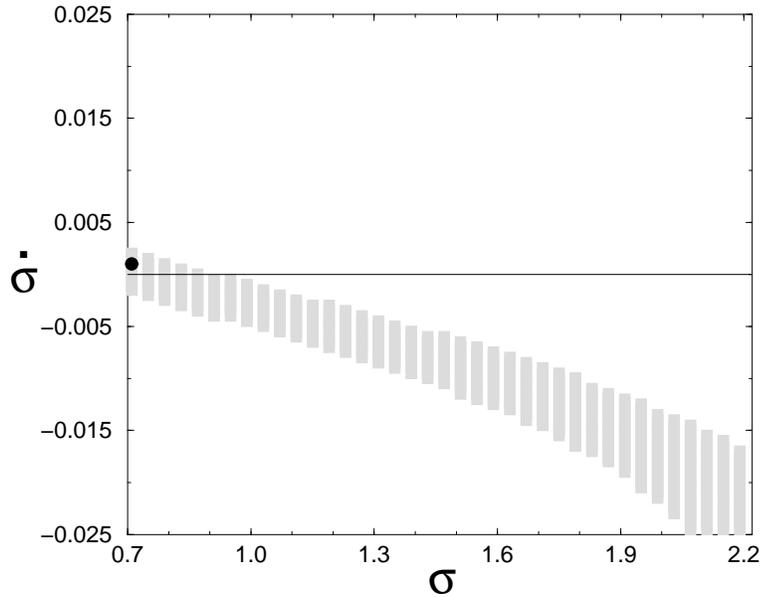}
\caption{The shaded area indicates the region of initial
conditions for which the system has a nonsingular evolution, and the dot
corresponds to the value actually chosen in the solution displayed in
Figs.~7-9. We have displayed only the part of the plane
with $\sigma>\sigma_{\rm sd}=\log 2$,
since modular invariance ensures that the figure is
invariant under the transformation 
$\sigma\rightarrow\sigma_{\rm sd} - \sigma$.}
\end{figure}

\begin{figure}
\label{fig11}
\centering
\includegraphics[width=0.6\linewidth,angle=-90]{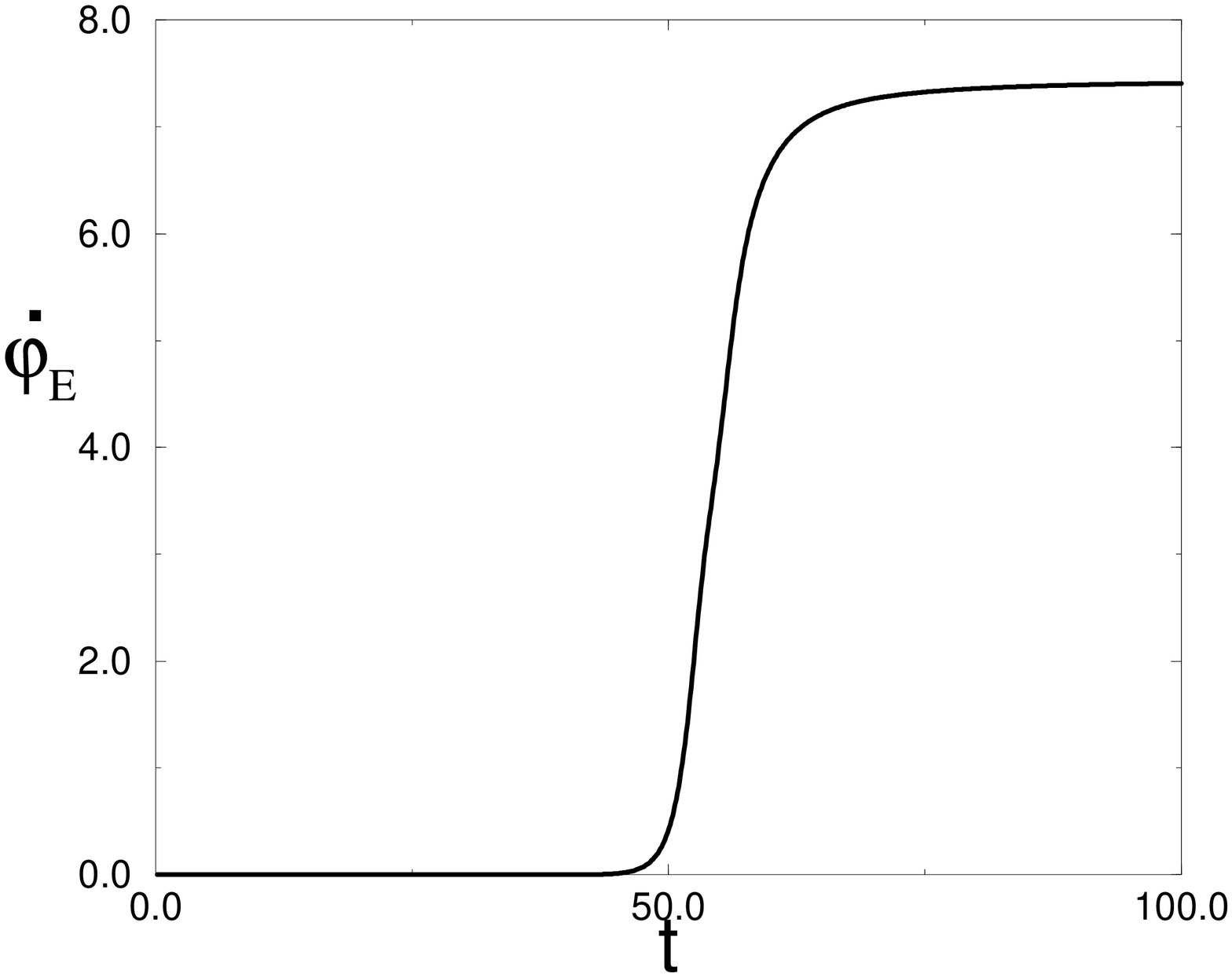}

\mbox{(a)}

\includegraphics[width=0.6\linewidth,angle=-90]{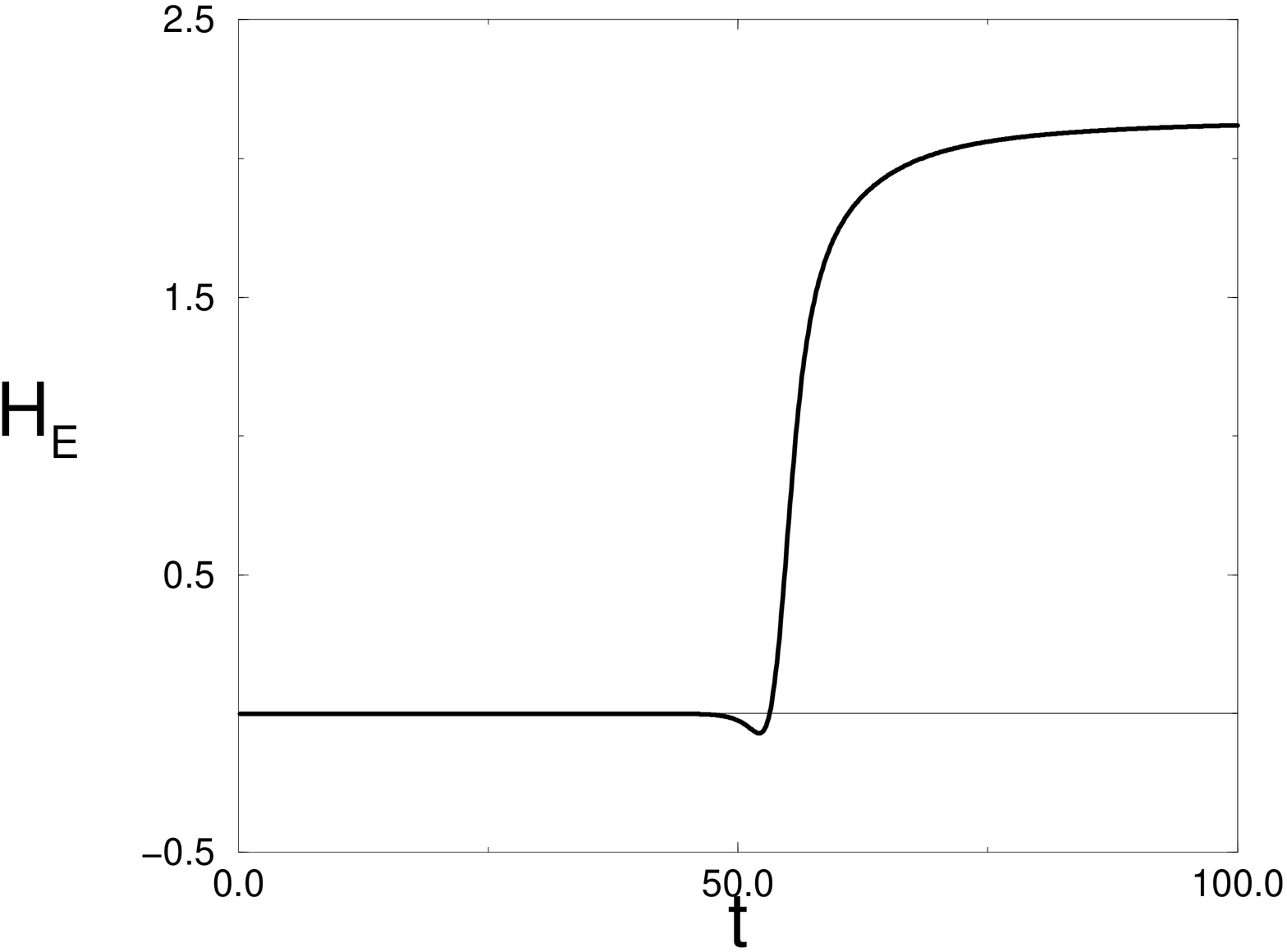}

\mbox{(b)}
\caption{(a) $\dvf_E$ and (b) $H_E$ against string time.}
\end{figure}
\clearpage

\section{Transition to a D-brane dominated regime}

We now  discuss the limitations on the validity of our
solutions. As it is clear from Fig.~9, at large values of time we are
deep into the strong coupling regime, $g^2\gg 1$. Can we still believe
our solutions? In our action we have included the
corrections to the \kal potential at {\em all} perturbative orders, while
other operators, like $R$ and $R_{\mu\nu\rho\sigma}^2$ are protected
by non renormalization theorems. Therefore, despite the ambiguities
that we have discussed for the four-derivative terms,  due in
particular  to naked $R_{\mu\nu}^2$ and $R^2$ terms, one might 
be tempted to argue
that the solution is at least representative of the behaviour at
strong coupling.  However, this point of view is untenable, and at
some point the perturbative approach itself breaks down.

To understand this point, it is useful to work in the Einstein frame.
The two-derivative part of our action then reads
\be
{\cal S}^{E}_2=\frac{1}{2\alpha '}\int d^4x \sqrt{-g} 
\left[
R-\frac{1}{2}Z_{\vf}\pa_{\mu}\varphi\pa^{\mu}\varphi
-\frac{3}{2}\pa_{\mu}\sigma\pa^{\mu}\sigma
\right]
\ee
with
\be
Z_{\vf}=1-2 e^{\vf}G(\varphi )=
1-2e^{\vf}
\left(\frac{3\kappa}{2}\right) \frac{6+\kappa
e^{\varphi}}{(3+\kappa e^{\varphi})^2}\, .
\ee
At weak coupling $e^{\vf}G(\vf )\ll 1$ and the kinetic term of the
dilaton has the `correct' sign. However, as $e^{\vf}\ra\infty$,
$e^{\vf}G(\vf )\ra 3/2$ and $Z_{\vf}<0$; $Z_{\vf}$ vanishes at
a critical value of $g^2=e^{\vf}$ given by
\be\label{067}
g_c^2= \frac{3}{2\kappa}\left(\sqrt{6}-2\right)\simeq\frac{0.67}{\kappa}\, .
\ee
At $e^{\vf}>g_c^2$ it appears that the dilaton becomes
ghost-like. The situation is quite similar to what happens in
Seiberg-Witten theory~\cite{SW}, and the physical interpretation
is the same. We can rescale the dilaton so that
it has a canonically normalized kinetic term
$(-1/2)\pa^{\mu}\vf\pa_{\mu}\vf$, and in terms of the rescaled 
dilaton the four-derivative
interactions, and in general all interactions involving the dilaton,
become strong as we approach $g_c$, and
formally  diverge at the critical point. This signals that
the effective action approach that we have used breaks down and
we must move to a new description, where the light degrees of freedom
are different. In the Seiberg-Witten model the new weakly coupled modes
are related to the original ones by an S-duality transformation.
In string theory the S-dual variables are given by D-branes:
for instance, in the Einstein frame, a fundamental string has a
tension
$g^{1/2}/(2\pi\alpha ')$ while a D-string has a tension
$g^{-1/2}/(2\pi\alpha ')$~\cite{Dbrane}.
This suggest that, if the cosmological 
evolution enters the regime $e^{\vf}>g^2_c$, the
effective action approach that we have used breaks down, and we enter
a new regime, which cannot be described in terms of a classical
evolution of massless modes of a closed string;  in this regime
we must  resort to a description in terms of D-branes.

More precisely, the condition $Z_{\vf}=0$ identifies the critical
point only if $\dvf, H$ can be neglected. In fact,
the equation of motion for $\vf$ in the Einstein frame reads 
(we insert for future use also a potential $V(\vf )$)
\be\label{fric}
M_\vf \ddot{\vf}_E = -3 A H_E\dvf_E
-V'\, ,\label{dfE}
\ee
where
\be 
M_{\vf}=\left[ 1-2e^\varphi G(\vf )-
3\Delta (\sigma ) \dvf_E^2\right]+\ldots\, ,
\ee
and
\be
A=1-2e^{\vf}G(\vf )-\Delta (\sigma )\dot{\vf}_E^2+\ldots\, .
\ee
The dots denote tree-level $\alpha '$ corrections (which are neglegible
at the later stage of the evolution).
We recall that we found regular solutions for
$\Delta <0$. So we see that, if we  include the effect of the term
$|\Delta |\dvf_E^2$,
the critical line is  given by the condition $M_{\vf}=0$ or
\be\label{curve}
Z_{\vf}+3|\Delta | \dvf_E^2=0\, ,
\ee
rather than
$Z_{\vf}=0$. Of course when $\dvf_E$ approaches one
we should at least include all higher
powers in the $\alpha '$ expansion. More importantly,
in the regime  where $H$ or 
$\dvf$ approach one 
other restrictions on the validity of the effective
action appear, and have been discussed in ref.~\cite{MR},
embedding the 10-dimensional theory into 11-dimensional M-theory
compactified on $S^1$. In Fig.~12 (adapted from ref.~\cite{MR})
on the vertical axis we show  $H$,
in the string frame. This is  an indicator of the curvature and
therefore of the typical energy
scale of the solution. One might as well use $\dvf$, but of course
precise numerical values here are not very important.
In this graph 
we prefer to use the string frame quantity
$H$ because in this case the $\alpha '$ corrections become
important when $H\sim 1$,
while in terms of $H_E$ this condition becomes
$e^{-\vf} H_E^2\sim 1$.

The solid line $H\sim 1/g$ separates the region
where an effective 10-dimensional description is possible, from the
truly 11-dimensional regime. The region just above the line labelled
11d-Sugra is described by 11-dimensional supergravity, while above the
line labelled Dp-branes we are in the full M-theory regime. For a
discussion of the crossover between these regimes we refer the reader
to ref.~\cite{MR}. Of course,  again, the  position of the line
separating the full M-theory regime from the 11-D supergravity regime
is only indicative, and we have arbitrarily chosen its position  so that
it meets the curve $H=1/g$ exactly at $H=g=1$.

On the 10-dimensional side we have also drawn as a solid line the
curve given by eq.~(\ref{curve}), which is another critical line where
a change of regime occurs. When $\dvf_E$ is not small,  the form of
this curve is only indicative. The position of $g_c$ depends on
$\kappa$, eq.~(\ref{067}), and the graph refers to a $Z_6$ orbifold, 
$\kappa\simeq 0.19$.
After the curve enters in the 11-dimensional
region it is not anymore meaningful.
The label `S-duality' means that,
crossing this line, we enter a regime
where the light degrees of freedom are related to the original ones by
S-duality. 
On the same figure we display the
solution of  Fig.~7a, labelled by the arrows.
The solution for $H$ will eventually decrease,  but this
only happens at very large values of $g^2$ (see Fig.~9), 
and we see that
the solution enters the 11-dimensional domain before it starts
decreasing. At this point, it looses its validity.

\begin{figure}
\label{fig12}
\centering
\includegraphics[width=0.6\linewidth,angle=-90]{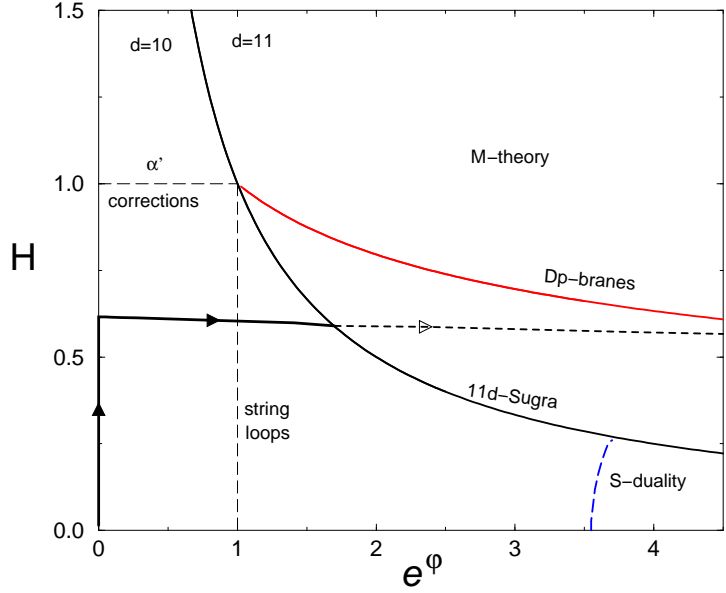}
\caption{The `phase diagram' of M-theory compactified on $S^1$.
See the text for explanations of the various lines. The cosmological
solution found in sect.~3 is marked by the arrows. }
\end{figure}

Finally, we found that it is not possible to stabilize
the dilaton in our solution at a
minimum of a  potential, as could be generated for instance by
gaugino condensation~\cite{VY}. In fact at the later stage of the
evolution the tree level $\alpha '$ corrections are neglegible, as we
see in Fig.~7a, and 
$M_{\vf}\simeq  1-2e^{\varphi} G(\vf )+3|\Delta | \dvf_E^2$. 
If we would stabilize $\vf$ around the minimum of the
potential, it should first oscillate around the minimum and at the inversion
points $\dvf_E =0$, so that here the coefficient of $\ddot{\vf}_E$ 
in eq.~(\ref{fric}) becomes
$\simeq Z_{\vf}$. As shown in Fig.~13, 
this quantity is negative after we cross the
$H_E>0$ line. As we discussed, this is not a problem for the consistency
of the solution as long as $\dvf$ is not small (in fact, Fig.~12 shows
that the limitation on the validity of the solution is rather given by the
crossing into the 11-dimensional region), but  it is clear that no
consistent solution with $\dvf_E =0$ can be obtained trying to stabilize
the dilaton with a potential. In fact, if we try to force
$\dvf_E$ to a small value, 
the coefficient of $\ddot{\vf}_E$ in eq.~(\ref{fric}) becomes approximately
equal to $Z_{\vf}$, which at this stage is negative. 
Therefore the evolution runs away from the minimum of the
potential.  Numerically, we have found  that, including a potential in the
numerical integration of the equations of motion, when the solution
approaches the minimum of the potential the numerical precision,
monitored by the constraint equation, degrades immediately and the
solution explodes.

Therefore, in our scenario, the problem of the dilaton stabilization
can only be solved after the solution enters in the 
non-perturbative regime.

\begin{figure}
\label{fig13}
\centering
\includegraphics[width=0.6\linewidth,angle=-90]{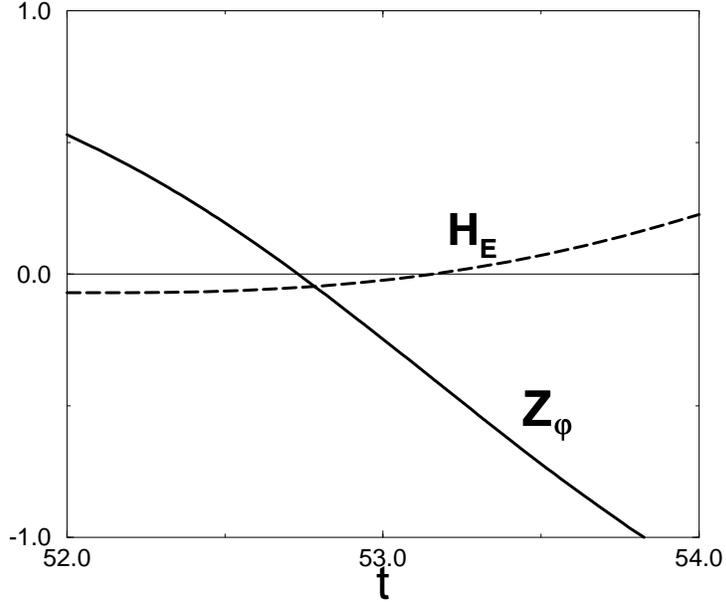}
\caption{The  evolution of 
$Z_{\vf}\equiv 1-2e^\varphi G(\vf )$ 
close to the point where $H_E$ becomes positive, against string frame 
time.}
\end{figure}

\clearpage

\section{Conclusions}
In this paper we have tried to penetrate into the strong coupling
regime of the cosmological evolution derived from string theory. 
This regime is crucial for an understanding of the
big-bang singularity in string theory, but since loop corrections do
not tame the growth of the coupling while remaining within the weak
coupling domain, it is clear that
a knowledge limited to, say, one-loop corrections is 
of little use, and we really need to have at least a glimpse into the
structure of the corrections at all perturbative orders. Luckily, 
for the effective action of orbifold compactifications of
heterotic string theory, supersymmetry
and modular invariance impose strong constraints on the form of the
corrections at all orders. In particular, the kinetic terms of the
dilaton is known exactly, while other operators, like $R$ and
$R_{\mu\nu\rho\sigma}^2$, are protected by non-renormalization
theorems. Therefore, in spite of some ambiguities in the choice of the
four-derivative terms, present both at tree level and for their loop
corrections, one can try to investigate string cosmology beyond the
weak coupling domain, and to obtain at least some indications of what
a well motivated stringy scenario looks like.  

As a first step, we have therefore tried to push this perturbative
approach as far as possible, following the evolution even in the
strong coupling domain $g\gg 1$. We have found solutions with
interesting properties, that in the string frame
start with a
pre-big-bang superinflationary phase,  go through a phase with $H,\dvf$
approximately constant and of order one in string units, 
(a phase  that replaces the big-bang singularity) and then match to a
regime with $H, \dvf$ decreasing. Apart from their intrinsic interest,
these solutions provide an illustration of the interplay between
$\alpha '$ and loop corrections in string cosmology, and give an
explicit realization of the general picture emerged from the 
works~[1,5-12].
Probably the main element that is missing from this part of
the  analysis is the
inclusion of non-local terms. These might model the backreaction due
to quantum particles production, which  might  play an important role in
the graceful exit transition~\cite{BMUV2}. Unfortunately, these 
are quite difficult to include in a numerical analysis.

Despite some nice properties, the cosmological model
that we have presented still
have some unsatisfactory features, and in particular the dilaton could
not be stabilized with a potential, and so this model
cannot be the end of the
story.
 
On the other hand we have found that, if we look at our solution from
the broader perspective of 11-dimensional theory, it ceases to be
valid as soon as we enter into the strong coupling region, even if
one includes perturbative corrections at all orders.
Thus, we think that our analysis reveals quite clearly the
direction that should be taken to make further progress. 
As already discussed in ref.~\cite{MR}, when we move
toward large curvatures we meet critical lines in the $(H,g)$ plane,
beyond which D-branes becomes the relevant degrees of
freedom. Here we have found another critical line at strong
coupling; beyond this line the light modes relevant for an effective
action approach are obtained by an
S-duality transformation and are therefore again naturally interpreted as
D-branes. 
The combination of these critical lines, shown in fig.~12,
and the behaviour of our solutions, also displayed on the same graph,
suggest that the evolution enters unavoidably  the regime where  new
descriptions set in. The understanding and the smoothing of the
big-bang singularity  therefore  requires the use of truly
non-perturbative string physics.

\vspace{5mm}
Acknowledgments. We thank 
Maurizio Gasperini, Ken Konishi,  Kostas Kounnas,  Toni Riotto, 
Gabriele Veneziano and Fabio Zwirner for useful discussions.
We are grateful to  Ramy Brustein for very useful comments on the
manuscript.

\end{document}